\documentclass[iop]{emulateapj}

\usepackage{natbib,aas_macros}
\usepackage{multirow,color}

\begin{document}
\shorttitle{
Disk-like Structures in High-$z$ Dusty Star-Forming Galaxy
}
\shortauthors{Fujimoto et al.}
\slugcomment{ApJ in press}

\title{%
{ALMA 26 arcmin$^{2}$ Survey of GOODS-S at One-millimeter (ASAGAO):} 
Average Morphology of High-$\lowercase{z}$ Dusty Star-Forming Galaxies is 
an Exponential-Disk ($\lowercase{n} \simeq 1$) 
}

\author{%
Seiji Fujimoto\altaffilmark{1}, 
Masami Ouchi\altaffilmark{1,2}, 
Kotaro Kohno\altaffilmark{3,4}, 
Yuki Yamaguchi\altaffilmark{3}, 
Bunyo Hatsukade\altaffilmark{3}, 
Yoshihiro Ueda\altaffilmark{5}, 
Takatoshi Shibuya\altaffilmark{1}, 
Shigeki Inoue\altaffilmark{2}, 
Taira Oogi\altaffilmark{2}, 
Sune Toft\altaffilmark{6}, 
Carlos G$\acute{\rm o}$mez-Guijarro\altaffilmark{6}, 
Tao Wang\altaffilmark{3,7}, 
Daniel Espada \altaffilmark{7,8}, 
Tohru Nagao\altaffilmark{9}, 
Ichi Tanaka\altaffilmark{10}, 
Yiping Ao\altaffilmark{7}, 
Hideki Umehata\altaffilmark{3,11,12}, 
Yoshiaki Taniguchi\altaffilmark{11}, 
Kouichiro Nakanishi \altaffilmark{7,8}, 
Wiphu Rujopakarn\altaffilmark{2, 13, 14}, 
R.\,J. Ivison\altaffilmark{15,16}, 
Wei-hao Wang\altaffilmark{17}, 
Minju M. Lee\altaffilmark{7,18}, 
Ken-ichi Tadaki\altaffilmark{7},
Yoichi Tamura\altaffilmark{19}, 
and J.\,S. Dunlop\altaffilmark{20} 
}

\email{sfseiji@icrr.u-tokyo.ac.jp}

\altaffiltext{1}{%
Institute for Cosmic Ray Research, The University of Tokyo,
Kashiwa, Chiba 277-8582, Japan
}
\altaffiltext{2}{%
Kavli Institute for the Physics andMathematics of the Universe 
(Kavli IPMU), WPI, The University of Tokyo, 
Kashiwa, Chiba 277-8583, Japan
}
\altaffiltext{3}{Institute of Astronomy, Graduate School of Science,
The University of Tokyo, 2-21-1 Osawa, Mitaka, Tokyo 181-0015, Japan
}
\altaffiltext{4}{Research Center for the Early Universe, Graduate
School of Science, The University of Tokyo, 7-3-1 Hongo, Bunkyo-ku,
Tokyo 113-0033, Japan
}
\altaffiltext{5}{%
Department of Astronomy, Kyoto University, Kyoto 606-8502, Japan
}
\altaffiltext{6}{%
Cosmic Dawn Center (DAWN), Niels Bohr Institute, University of Copenhagen, Denmark
}
\altaffiltext{7}{%
National Astronomical Observatory of Japan, Mitaka, 
Tokyo 181-8588, Japan
}
\altaffiltext{8}{%
Department of Astronomical Science, SOKENDAI
(The Graduate University of Advanced Studies),
2-21-1 Osawa, Mitaka, Tokyo 181-8588, Japan 
}
\altaffiltext{9}{%
Research Center for Space and Cosmic Evolution, Ehime University, 
2-5 Bunkyo-cho, Matsuyama, Ehime 790-8577, Japan
}
\altaffiltext{10}{%
Subaru Telescope, National Astronomical Observatory of Japan, 
650 North A’ohoku Place, Hilo, HI 96720, USA
}
\altaffiltext{11}{%
The Open University of Japan, 2-11, Wakaba, Mihama-ku, Chiba, Chiba 261-8586, Japan
}

\altaffiltext{12}{%
RIKEN Cluster for Pioneering Research, 2-1 Hirosawa, Wako-shi, Saitama 351-0198, Japan
}
\altaffiltext{13}{%
Department of Physics, Faculty of Science, Chulalongkorn University, 
254 Phayathai Road, Pathumwan, Bangkok 10330, Thailand
}
\altaffiltext{14}{%
National Astronomical Research Institute of Thailand (Public Organization), 
Don Kaeo, Mae Rim, Chiang Mai 50180, Thailand
}
\altaffiltext{15}{%
Institute for Astronomy, University of Edinburgh, Royal Observatory, 
Blackford Hill, Edinburgh EH9 3HJ, UK
}
\altaffiltext{16}{%
European Southern Observatory, 
Karl Schwarzschild Str. 2, D-85748 Garching, Germany
}
\altaffiltext{17}{%
Institute of Astronomy and Astrophysics, Academia Sinica, Taipei 10617, Taiwan
}
\altaffiltext{18}{%
Department of Astronomy, The University of Tokyo, 7-3-1 Hongo, Bunkyo-ku, Tokyo 133-0033, Japan
}
\altaffiltext{19}{%
Department of Physics, Nagoya University, Furo-cho, Chikusa-ku, Nagoya 464-8601, Japan
}
\altaffiltext{20}{%
Institute for Astronomy, University of Edinburgh, Royal Observatory, Blackford Hill, Edinburgh EH9 3HJ, UK
}

\def\aj{AJ}%
\def\actaa{Acta Astron.}%
\def\araa{ARA\&A}%
\def\apj{ApJ}%
\def\apjl{ApJ}%
\def\apjs{ApJS}%
\def\ao{Appl.~Opt.}%
\def\apss{Ap\&SS}%
\def\aap{A\&A}%
\def\aapr{A\&A~Rev.}%
\def\aaps{A\&AS}%
\def\azh{AZh}%
\def\baas{BAAS}%
\def\bac{Bull. astr. Inst. Czechosl.}%
\def\caa{Chinese Astron. Astrophys.}%
\def\cjaa{Chinese J. Astron. Astrophys.}%
\def\icarus{Icarus}%
\def\jcap{J. Cosmology Astropart. Phys.}%
\def\jrasc{JRASC}%
\def\mnras{MNRAS}%
\def\memras{MmRAS}%
\def\na{New A}%
\def\nar{New A Rev.}%
\def\pasa{PASA}%
\def\pra{Phys.~Rev.~A}%
\def\prb{Phys.~Rev.~B}%
\def\prc{Phys.~Rev.~C}%
\def\prd{Phys.~Rev.~D}%
\def\pre{Phys.~Rev.~E}%
\def\prl{Phys.~Rev.~Lett.}%
\def\pasp{PASP}%
\def\pasj{PASJ}%
\def\qjras{QJRAS}%
\def\rmxaa{Rev. Mexicana Astron. Astrofis.}%
\def\skytel{S\&T}%
\def\solphys{Sol.~Phys.}%
\def\sovast{Soviet~Ast.}%
\def\ssr{Space~Sci.~Rev.}%
\def\zap{ZAp}%
\def\nat{Nature}%
\def\iaucirc{IAU~Circ.}%
\def\aplett{Astrophys.~Lett.}%
\def\apspr{Astrophys.~Space~Phys.~Res.}%
\def\bain{Bull.~Astron.~Inst.~Netherlands}%
\def\fcp{Fund.~Cosmic~Phys.}%
\def\gca{Geochim.~Cosmochim.~Acta}%
\def\grl{Geophys.~Res.~Lett.}%
\def\jcp{J.~Chem.~Phys.}%
\def\jgr{J.~Geophys.~Res.}%
\def\jqsrt{J.~Quant.~Spec.~Radiat.~Transf.}%
\def\memsai{Mem.~Soc.~Astron.~Italiana}%
\def\nphysa{Nucl.~Phys.~A}%
\def\physrep{Phys.~Rep.}%
\def\physscr{Phys.~Scr}%
\def\planss{Planet.~Space~Sci.}%
\def\procspie{Proc.~SPIE}%
         
\def\tcr{\textcolor{red}}
\def\tcb{\textcolor{black}}
\def\tcm{\textcolor{black}}
\def\tck{\textcolor{black}}

\def\rme{\rm e}
\def\rmFIR{\rm FIR}
\def\rmIR{\rm IR}
\def\rmopt{\rm opt}
\def\itHubble{\it Hubble}
\def\rmyr{\rm yr}

\begin{abstract}
We present morphological properties of dusty star-forming galaxies at $z=$1$-$3 
determined with the high-resolution (FWHM$\sim$$0\farcs19$) Atacama Large Milllimeter/submilimeter Array (ALMA) 1-mm map of our ASAGAO survey covering a 
26-arcmin$^{2}$ area in GOODS-S. 
In conjunction with the ALMA archival data, 
our sample consists of 45 ALMA sources with infrared luminosity $L_{\rmIR}$ range of $\sim10^{11}-10^{13}\,L_{\odot}$. 
To obtain an average rest-frame far-infrared (FIR) profile, we perform individual measurements and careful stacking of the ALMA sources using the $uv$-visibility 
method that includes positional-uncertainty and smoothing-effect evaluations through Monte-Carlo simulations. 
We find that our sample has the average FIR-wavelength S$\acute{\rme}$rsic index and effective radius of $n_{\rmFIR}=1.2\pm0.2$ and $R_{\rm e,FIR}=$1.0$-$1.3 kpc, 
respectively, additionally with a point-source component at the center, indicative of the existence of AGN. 
The average FIR profile agrees with a morphology of an exponential-disk clearly distinguished from a deVaucouleurs-spheroidal profile (S$\acute{\rme}$rsic index of 4). 
We also examine the rest-frame optical S$\acute{\rme}$rsic index $n_{\rmopt}$ and effective radius $R_{\rme,opt}$ with deep {\itHubble\,\,Space\,\,Telescope} (HST) images. 
Interestingly, we obtain $n_{\rmopt}=0.9\pm0.3$ ($\simeq$$n_{\rmFIR}$) and  $R_{\rme,opt}=3.2\pm0.6$ kpc ($>R_{\rme,FIR}$), 
suggesting that the dusty disk-like structure is embedded within a larger stellar disk. 
The rest-frame UV and FIR data of HST and ALMA provide us a radial profile of the total star-formation rate (SFR), 
where the infrared SFR dominates over the UV SFR at the center. Under the assumption of a constant SFR, a compact stellar distribution in $z\sim$1$-$2 compact quiescent galaxies (cQGs) is well reproduced, while a spheroidal stellar morphology of cQGs ($n_{\rmopt}=4$) cannot, suggestive of other important mechanism(s) such as dynamical dissipation.
\end{abstract}

\keywords{%
galaxies: formation ---
galaxies: evolution ---
galaxies: high-redshift 
}

\section{Introduction}
\label{sec:intro}
{ The size and morphology of} high-redshift galaxies provide us invaluable insights into galaxy evolution and formation. 
The S$\acute{\rm e}$rsic index $n$ \citep{sersic1963,sersic1968} and the effective radius $R_{\rm e}$ are key quantities to evaluate the size and morphological properties. 

In the rest-frame ultra-violet (UV) {to} optical wavelengths, 
the {\it Hubble\,Space\,Telescope} (HST) has revealed {the $n$ and $R_{\rm e}$ properties for the high-redshift galaxies up to $z\sim6$ and even $z\sim10$, respectively \citep[e.g.,][]{shen2003, ferguson2004, hathi2008, oesch2010, ono2013, vanderwel2014, shibuya2015, bowler2017, kawamata2015,kawamata2018}.} 
These HST studies show that star-forming and quiescent galaxies have an exponential-disk morphology with a S$\acute{\rm e}$rsic index in the rest-frame optical wavelength $n_{\rm opt} \sim1$ and a spheroidal morphology with $n_{\rm opt}\sim4$, respectively. 
Moreover, the quiescent galaxies are more compact than the star-forming galaxies at a given stellar mass. 
These HST results indicate that the transition from star forming to quiescent galaxy may be associated with transformation in both size and morphology.
However, the rest-frame UV and optical studies cannot unveil the actively star-forming regions obscured by dust. 
Due to an extreme star formation rates in these dusty star-forming galaxies, the stars formed may eventually dominate over the stars already present in the host galaxies. 
Studies of the morphology and size in the rest-frame far-infrared (FIR) wavelength are thus important to comprehensively understand the evolutionary connections between the high-$z$ star-forming and compact quiescent galaxies. 

The Atacama Large Millimeter/submillimeter Array (ALMA) enables us to measure $n$ and $R_{\rm e}$ in the rest-frame FIR wavelength, $n_{\rm FIR}$ and $R_{\rm e,FIR}$, due to its high sensitivity and angular resolution. 
For bright submillimeter galaxies (SMGs; $S_{\rm 1mm}\gtrsim1$ mJy), 
recent ALMA studies report that SMGs have the exponential-disk morphology with $n_{\rm FIR}=0.9\pm0.2$ \citep{hodge2016}, 
and $R_{\rm e,FIR}$ of $\sim1-2$ kpc \citep[e.g.,][]{ikarashi2015,simpson2015a,hodge2016,fujimoto2017} smaller than $R_{\rm e}$ in the rest-frame optical wavelength $R_{\rm e,opt}$ of {$\sim3-4$ kpc \citep[e.g.,][]{targett2011,targett2013,chen2015,fujimoto2017}}. 
For faint SMGs ($S_{\rm 1mm}<1$ mJy), although there are several attempts to estimate $R_{\rm e,FIR}$ by using deep ALMA imagings \citep{rujopakarn2016,gonzalez2017,fujimoto2017}, 
large uncertainties still remain due to the small number statistics and observational challenges. 

There are two major observational challenges to measure $n_{\rm FIR}$ and $R_{\rm e,FIR}$ for faint SMGs. 
First is the sensitivity. To perform a secure profile fitting, high signal-to-noise (S/N) level is required.  
Extremely deep observations are thus needed. 
Second is the angular resolution. 
\cite{fujimoto2017} report a positive correlation between the FIR size and luminosity, 
which suggests that faint SMGs are more compact than bright SMGs. 
To resolve the compact objects, we need high angular resolution. 

In this paper, we perform $n_{\rm FIR}$ and $R_{\rm e,FIR}$ measurements for faint SMGs via the stacking technique on the $uv$-visibility plane, 
utilizing ALMA sources identified in the high-resolution ($\sim$$0\farcs19$) ALMA 1mm survey of ASAGAO: ALMA twenty-Six Arcmin$^{2}$ survey of GOODS-S at One-millmeter (PI: K. Kohno). 
In conjunction with individual bright SMG results from the archive, we study the average morphology of high-$z$ dusty star-forming galaxies. 
The structure of this paper is as follows. 
In Section 2, the observations and the data reduction are described.  
Section 3 outlines the method of source detections, flux density and position measurements, multi-wavelength properties of our sample, the stacking process, and the $n_{\rm FIR}$ and $R_{\rm e,FIR}$ measurements. 
We report the results for $n_{\rm FIR}$, $R_{\rm e,FIR}$, $n_{\rm opt}$, $R_{\rm e,opt}$, and morphological classification for the dusty star-forming galaxies as a function of $L_{\rm IR}$ in Section 4. 
In Section 5, we discuss {possible origins of a compact component at the center and evolutions of the size and morphology from star-forming to compact quiescent galaxies.}
A summary of this study is presented in Section 6. 

Throughout this paper, we assume a flat universe with 
$\Omega_{\rm m} = 0.3$, 
$\Omega_\Lambda = 0.7$, 
$\sigma_8 = 0.8$, 
and $H_0 = 70$ km s$^{-1}$ Mpc$^{-1}$. 
We use magnitudes in the AB system \cite{oke1983}. 
{We perform the morphology diagnostic basically based on the S${\rm e}$rsic index $n$ with a single S$\acute{\rm e}$rsic profile, 
while we examine the detail morphological classification with deep HST/$H$-band images in Section \ref{sec:opt_morph}. 
We adopt the $R_{\rm e}$ values that are circularized by 
\begin{eqnarray}
R_{\rm e} = r_{\rm e, maj} \times \sqrt{q}  \, , 
\end{eqnarray}
where $r_{\rm e, maj}$ and $q$ are the half-light radius along the semi-major axis and the axis ratio, respectively. 
}

\section{Data and Reduction} 
\label{sec:data}

In September 2016, the ASAGAO survey carried out ALMA Band 6 observations over an area of $\sim$26 arcmin$^{2}$ 
in The Great Observatories Origins Deep Survey South (GOODS-S; \citealt{vanzella2005}) 
with a total observing time of 45 hours (ALMA project ID: 2015.1.00098.S, PI: K. Kohno; see also Ueda et al. 2017). 
The array configuration was C40-6 with 38$-$45 antennas whose baseline length takes a range of 15$-$3247 m.  
The frequency setting was such that in the two tunings centered at 1.14 mm and 1.18mm, 
where the frequency coverages were 244$-$248 GHz, 253$-$257 GHz, 259$-$263 GHz, and 268$-$272 GHz. 

The data were reduced with the Common Astronomy Software Applications package (CASA; \citealt{mcmullin2007}).  
The details of the data calibration and reduction are described in Hatsukade et al. (in prep.). 
The entire map was produced by the CLEAN algorithm with the {\sc tclean} task. 
The CLEAN boxes were set at the peak pixel positions with S/N $\geq$ 5 in the manual mode. 
For the CLEAN box, we adopt a circle with a radius of $1\farcs0$.   
The CLEAN routines were proceeded down to the 2$\sigma$ level. 
The final natural-weighted image is characterized by a synthesized beam size of $0\farcs21\times0\farcs17$ and an rms noise level of 38 $\mu$Jy/beam. 
We also produced another map with a $uv$-taper of 160 k$\lambda$ whose final synthesized beam size is $0\farcs94\times0\farcs67$ and rms noise level is 87 $\mu$Jy/beam. 
We refer to the natural-weighted (high-resolution) and the $uv$-tapered (low-resolution) images as "HR" and "LR" maps, respectively. 
{In the following analyses, we use the HR map except for source detections and positional measurements in Section \ref{sec:source_detection} and Section \ref{sec:fp_measure}.} 

\section{Data Analysis}
\label{sec:data_analysis}

\begin{table*}
{\scriptsize
\caption{Our ASAGAO Source Catalog}
\begin{tabular}{llllllllll}
\hline
\hline
 ID(ASAGAO) & ID(ZFOURGE) & R.A.(ALMA) & Dec.(ALMA) & S/N & $S_{\rm total}$ & $z_{\rm phot}$ ($z_{\rm spec}$) & $\log(L_{\rm IR})$ & $\log(M_{\rm star})$  & $\log({\rm SFR})$\\
                       &                          & (J2000) & (J2000)                 &        &   (mJy)              &                                                     &     ($L_{\odot}$)  &    ($M_{\rm star}$)  & ($M_{\odot}/$yr) \\
 (1)                  &   (2)                  &    \multicolumn{2}{c}{(3)}      &  (4)  &     (5)                &    			(6)                      &           (7)             &              (8)          &  (9)    \\
\hline
1 & 17856 & 53.118798 & -27.782886 & 20.27 & 2.33$\pm$0.11 & 2.38 ($-$) & 12.76$_{-0.0}^{+0.0}$ & 11.44$_{-0.0}^{+0.0}$ & 2.7$_{-0.0}^{+0.0}$ \\
2 & 13086 & 53.148854 & -27.821189 & 19.82 & 2.04$\pm$0.1 & 2.58 (2.582) & 12.86$_{-0.03}^{+0.0}$ & 11.69$_{-0.0}^{+0.01}$ & 2.8$_{-0.02}^{+0.0}$ \\
3 & 18658 & 53.183412 & -27.776462 & 3.94 & 1.46$\pm$0.12 & 2.83 ($-$) & 12.75$_{-0.0}^{+0.0}$ & 10.91$_{-0.0}^{+0.0}$ & 2.46$_{-0.0}^{+0.0}$ \\
4 & 22177 & 53.198346 & -27.747876 & 12.01 & 0.89$\pm$0.08 & 1.93 ($-$) & 12.61$_{-0.0}^{+0.01}$ & 11.34$_{-0.0}^{+0.02}$ & 2.55$_{-0.0}^{+0.0}$ \\
5 & 18645 & 53.181375 & -27.777557 & 10.63 & 1.15$\pm$0.12 & 2.92 ($-$) & 12.5$_{-0.0}^{+0.0}$ & 11.52$_{-0.0}^{+0.0}$ & 2.38$_{-0.0}^{+0.0}$ \\
6 & 20298 & 53.137349 & -27.761634 & 3.90 & 1.17$\pm$0.13 & 0.52 (0.523) & 11.0$_{-0.0}^{+0.0}$ & 10.92$_{-0.0}^{+0.0}$ & 0.62$_{-0.0}^{+0.0}$ \\
7 & 21730 & 53.121846 & -27.752782 & 8.54 & 0.79$\pm$0.1 & 2.01 ($-$) & 12.18$_{-0.03}^{+0.0}$ & 11.42$_{-0.01}^{+0.01}$ & 2.01$_{-0.05}^{+0.0}$ \\
8 & 18701 & 53.160607 & -27.776218 & 7.94 & 0.65$\pm$0.08 & 2.61 ($-$) & 12.75$_{-0.0}^{+0.0}$ & 10.49$_{-0.0}^{+0.0}$ & 2.58$_{-0.0}^{+0.0}$ \\
9 & 21234 & 53.196561 & -27.757043 & 7.36 & 0.45$\pm$0.07 & 2.46 ($-$) & 11.78$_{-0.0}^{+0.0}$ & 10.19$_{-0.0}^{+0.0}$ & 1.28$_{-0.0}^{+0.0}$ \\
10 & 19033 & 53.131124 & -27.773185 & 7.32 & 0.64$\pm$0.1 & 2.22 (2.225) & 12.42$_{-0.0}^{+0.0}$ & 11.67$_{-0.0}^{+0.0}$ & 2.42$_{-0.0}^{+0.0}$ \\
11 & 14580 & 53.185853 & -27.810041 & 6.40 & 0.57$\pm$0.1 & 2.81 (2.593) & 11.97$_{-0.0}^{+0.0}$ & 10.97$_{-0.0}^{+0.0}$ & 2.02$_{-0.0}^{+0.0}$ \\
14 & 22760 & 53.199582 & -27.742692 & 6.36 & 2.93$\pm$0.5 & 2.16 ($-$) & 12.46$_{-0.0}^{+0.0}$ & 11.13$_{-0.0}^{+0.0}$ & 2.06$_{-0.0}^{+0.0}$ \\
26 & 17733 & 53.143495 & -27.783282 & 5.61 & 0.91$\pm$0.18 & 1.62 ($-$) & 12.12$_{-0.02}^{+0.0}$ & 11.64$_{-0.0}^{+0.0}$ & 1.88$_{-0.02}^{+0.0}$ \\
30 & 14122 & 53.119142 & -27.814017 & 4.93 & 0.66$\pm$0.14 & 3.32 ($-$) & 12.16$_{-0.25}^{+0.23}$ & 10.98$_{-0.08}^{+0.07}$ & 2.0$_{-0.26}^{+0.26}$ \\
42 & 19752 & 53.168622 & -27.770091 & 4.60 & 0.46$\pm$0.1 & 2.88 ($-$) & 9.27$_{-0.18}^{+0.62}$ & 9.13$_{-0.2}^{+0.0}$ & 0.43$_{-0.02}^{+0.06}$ \\
52 & 12763 & 53.169797 & -27.824018 & 4.57 & 1.23$\pm$0.27 & 2.13 (2.13) & 12.18$_{-0.0}^{+0.0}$ & 11.18$_{-0.0}^{+0.0}$ & 2.23$_{-0.0}^{+0.0}$ \\
60 & 20728 & 53.125299 & -27.75958 & 3.82 & 0.65$\pm$0.15 & 0.64 (0.647) & 10.82$_{-0.07}^{+0.09}$ & 10.26$_{-0.05}^{+0.01}$ & 0.97$_{-0.03}^{+0.06}$ \\
66 & 14700 & 53.120107 & -27.808343 & 4.44 & 0.56$\pm$0.13 & 1.83 ($-$) & 12.15$_{-0.0}^{+0.0}$ & 11.78$_{-0.0}^{+0.0}$ & 1.9$_{-0.0}^{+0.0}$ \\
67 & 20694 & 53.16361 & -27.759021 & 3.80 & 0.66$\pm$0.15 & 1.06 ($-$) & 11.5$_{-0.0}^{+0.14}$ & 11.42$_{-0.0}^{+0.04}$ & 1.31$_{-0.0}^{+0.1}$ \\
72 & 16039 & 53.176213 & -27.796212 & 3.79 & 0.55$\pm$0.13 & 0.95 (0.996) & 11.49$_{-0.06}^{+0.0}$ & 10.91$_{-0.0}^{+0.06}$ & 1.59$_{-0.05}^{+0.0}$ \\
73 & 15702 & 53.166917 & -27.79882 & 4.36 & 0.76$\pm$0.17 & 1.93 (1.998) & 12.47$_{-0.0}^{+0.0}$ & 11.3$_{-0.0}^{+0.0}$ & 2.44$_{-0.0}^{+0.0}$ \\
90 & 19487 & 53.165582 & -27.76987 & 3.71 & 0.35$\pm$0.08 & 1.61 ($-$) & 11.98$_{-0.0}^{+0.0}$ & 11.61$_{-0.0}^{+0.0}$ & 1.78$_{-0.0}^{+0.0}$ \\
102 & 21414 & 53.202184 & -27.754991 & 3.69 & 0.47$\pm$0.11 & 2.18 ($-$) & 11.63$_{-0.14}^{+0.02}$ & 10.77$_{-0.02}^{+0.02}$ & 1.6$_{-0.15}^{+0.01}$ \\
103$\dagger$ & 14146 & 53.182012 & -27.814195 & 3.51 & 0.32$\pm$0.07 & 9.49 ($-$) & $-$ & $-$ & $-$ \\
113 & 13714 & 53.141667 & -27.816646 & 4.30 & 0.81$\pm$0.19 & 2.53 ($-$) & 12.11$_{-0.04}^{+0.06}$ & 11.43$_{-0.0}^{+0.05}$ & 1.98$_{-0.09}^{+0.02}$ \\
129 & 12998 & 53.120801 & -27.819049 & 3.66 & 0.41$\pm$0.09 & 1.09 (1.094) & 11.7$_{-0.05}^{+0.0}$ & 11.61$_{-0.0}^{+0.18}$ & 1.65$_{-0.15}^{+0.0}$ \\
132 & 18912 & 53.170922 & -27.775466 & 4.23 & 0.3$\pm$0.07 & 2.36 ($-$) & 11.9$_{-0.02}^{+0.02}$ & 11.09$_{-0.02}^{+0.04}$ & 1.78$_{-0.04}^{+0.02}$ \\
148 & 14419 & 53.161443 & -27.811162 & 4.23 & 0.35$\pm$0.08 & 2.77 ($-$) & 12.07$_{-0.0}^{+0.0}$ & 11.36$_{-0.0}^{+0.0}$ & 1.92$_{-0.0}^{+0.0}$ \\
159 & 19453 & 53.114953 & -27.767634 & 3.65 & 0.47$\pm$0.11 & 0.64 (0.67) & 11.36$_{-0.0}^{+0.02}$ & 11.41$_{-0.0}^{+0.0}$ & 1.07$_{-0.03}^{+0.0}$ \\
180 & 18813 & 53.151876 & -27.775504 & 4.21 & 0.31$\pm$0.07 & 1.05 (1.047) & 11.54$_{-0.0}^{+0.0}$ & 10.75$_{-0.0}^{+0.0}$ & 1.56$_{-0.0}^{+0.0}$ \\
235 & 18379 & 53.156533 & -27.779731 & 4.21 & 0.3$\pm$0.07 & 1.27 ($-$) & 10.51$_{-0.29}^{+0.0}$ & 9.19$_{-0.0}^{+0.07}$ & 0.53$_{-0.17}^{+0.0}$ \\
251 & 13325 & 53.143052 & -27.820783 & 4.14 & 0.51$\pm$0.13 & 1.60 ($-$) & 10.0$_{-0.17}^{+0.14}$ & 9.39$_{-0.04}^{+0.0}$ & 0.44$_{-0.0}^{+0.08}$ \\
260 & 16977 & 53.15376 & -27.790687 & 3.64 & 0.27$\pm$0.07 & 1.32 (1.318) & 11.49$_{-0.0}^{+0.0}$ & 10.23$_{-0.0}^{+0.0}$ & 1.5$_{-0.0}^{+0.0}$ \\
322 & 16952 & 53.15405 & -27.790933 & 4.10 & 0.29$\pm$0.06 & 1.88 ($-$) & 11.9$_{-0.0}^{+0.0}$ & 11.12$_{-0.0}^{+0.0}$ & 1.83$_{-0.0}^{+0.0}$ \\
357 & 22905 & 53.119885 & -27.743161 & 4.10 & 0.46$\pm$0.26 & 3.85 ($-$) & 12.84$_{-0.0}^{+0.03}$ & 11.74$_{-0.0}^{+0.13}$ & 2.76$_{-0.01}^{+0.0}$ \\
393 & 15467 & 53.178708 & -27.802711 & 4.06 & 0.27$\pm$0.07 & 2.74 ($-$) & 12.05$_{-0.06}^{+0.07}$ & 11.47$_{-0.07}^{+0.03}$ & 1.79$_{-0.1}^{+0.04}$ \\
399 & 18476 & 53.145257 & -27.777982 & 3.55 & 0.29$\pm$0.07 & 1.03 (1.097) & 11.52$_{-0.08}^{+0.0}$ & 10.91$_{-0.05}^{+0.0}$ & 1.46$_{-0.08}^{+0.0}$ \\
408 & 21043 & 53.127397 & -27.755151 & 3.54 & 0.29$\pm$0.08 & 0.66 (0.681) & 11.21$_{-0.02}^{+0.08}$ & 10.96$_{-0.08}^{+0.02}$ & 1.36$_{-0.01}^{+0.06}$ \\
570 & 20880 & 53.187413 & -27.760581 & 4.02 & 0.29$\pm$0.07 & 1.77 ($-$) & 10.52$_{-0.41}^{+0.6}$ & 8.71$_{-0.08}^{+0.17}$ & 0.55$_{-0.21}^{+0.4}$ \\
577 & 13876 & 53.131372 & -27.815024 & 3.54 & 0.32$\pm$0.07 & 1.70 ($-$) & 11.78$_{-0.0}^{+0.0}$ & 11.21$_{-0.0}^{+0.0}$ & 1.7$_{-0.0}^{+0.0}$ \\
584 & 19133 & 53.200006 & -27.774142 & 4.01 & 0.41$\pm$0.07 & 4.36 ($-$) & 12.93$_{-0.0}^{+0.05}$ & 11.51$_{-0.06}^{+0.0}$ & 2.88$_{-0.0}^{+0.06}$ \\
613 & 16449 & 53.135186 & -27.795602 & 3.97 & 0.27$\pm$0.07 & 1.89 ($-$) & 10.89$_{-0.6}^{+0.37}$ & 8.84$_{-0.08}^{+0.2}$ & 0.79$_{-0.38}^{+0.27}$ \\ \hline
\end{tabular}
\tablecomments{
\footnotesize{
(1) ASAGAO ID. 
(2) ZFOURGE ID. 
(3) ASAGAO source center in our ALMA map. 
(4) ALMA Peak S/N. 
(5) Spatially integrated ALMA flux density. 
(6) Photometric (spectroscopic) redshift in ZFOURGE catalog \citep{straatman2016}. 
(7) Stellar mass obtained by MAGPHYS. 
(8) IR luminosity obtained by MAGPHYS. 
(9) UV + IR SFR obtained by MAGPHYS.
\label{tab:source_catalog}}}
$\dagger$ Because the photometric redshift is too large probably due to poor constraints in the SED fitting, we do not use ID103 for our analyses in this paper. 
}
\end{table*}

\begin{table}
{\scriptsize
\caption{Additional Source Catalog from Archive}
\begin{tabular}{lllllllc}
\hline
\hline
ID   & R.A. & Dec.           & S/N  & $z_{\rm phot}$ & $\log(L_{\rm IR})$ & Ref.     \\
            &    \multicolumn{2}{c}{(J2000)}  &             &                & ($L_{\odot}$)  &  \\
 (1)       &    \multicolumn{2}{c}{(2)}          &  (3)      &       (4)     &           (5)         &  (6)\\
\hline
105           & 34.568703 & -4.919106    & 15.3  &  2.30   & 12.90$_{-0.07}^{+0.06}$ & 1   \\  
254 		& 53.203598 & -27.52034 & 19.8  &  2.72   & 12.70$_{-0.57}^{+0.14}$    & 2 \\ 
255		& 53.389057 & -27.99159   & 26.0  &  2.67 & 12.64$_{-0.33}^{+0.23}$    & 2  \\  
256 		& 53.030369 & -27.85580 & 43.3  &  2.12   & 12.60$_{-0.26}^{+0.20}$    & 2  \\ 
260 		& 52.937637 & -27.57688   & 26.7  &  2.33  & 12.31$_{-0.13}^{+0.35}$   & 2 \\ 
580           & 34.418846 & -5.219619   & 28.4  & 2.87   & 12.48$_{-0.03}^{+0.03}$   & 1 \\ 
581		& 34.421337 & -5.220879   & 16.2   & 2.90  &  12.33$_{-0.06}^{+0.05}$  & 1 \\ 
592           & 34.422413 & -5.181027  & 21.4   & 2.72   & 12.51$_{-0.05}^{+0.04}$   & 1 \\ 
648           & 53.118809  & -27.78287 & 49.0  &  2.31   & 12.84$_{-0.10}^{+0.10}$   &  3 \\  
649           & 53.020367  & -27.77991 & 21.0  &  2.01   & 12.78$_{-0.05}^{+0.04}$   &  3 \\     
651           & 53.137592  & -27.70021 & 16.4  &  2.45   & 12.45$_{-0.10}^{+0.10}$   &  3 \\  
653           & 53.148853 & -27.82118  & 39.9  & 2.58    & 12.83$_{-0.10}^{+0.10}$   &  3  \\ \hline 
\end{tabular}
\tablecomments{
\footnotesize{
(1) ALMA source ID presented in \cite{fujimoto2017}. 
(2) ALMA source center. 
(3) ALMA Peak S/N. 
(4) Photometric redshift. 
(5) IR luminosity. 
(6) Reference of the $L_{\rm IR}$ measurement ([1]: \citealt{fujimoto2017}; [2] \citealt{dacunha2015}; [3] \citealt{barro2016}). 
\label{tab:additional_catalog}}}
}
\end{table}

\subsection{Source Detection and Catalog}
\label{sec:source_detection}
We conduct source extractions for our ASAGAO map before primary beam correction. 
We select the peak pixel values above 3.5$\sigma$ levels. 
Here we use the LR map to improve the surface brightness sensitivity, 
because previous ALMA high-resolution studies show that the dust emission from the high-$z$ star-forming galaxies are well resolved with the angular resolutions similar to the HR map \citep[e.g.,][]{hodge2016,tadaki2017a}. 
We obtain 631 source candidates with the peak above the 3.5$\sigma$ levels. 

To identify reliable ALMA sources from the 631 source candidates, we carry out the following two steps. 
First, we select the ALMA sources that have optical to near-infrared (NIR) objects within a radius of $0\farcs5$ in the ZFOURGE catalog \citep{straatman2016}. 
If the ALMA source is largely extended, we adopt a larger radius up to the half-light radius of the ALMA source. 
Note that recent ALMA studies show that there is a systematic astrometric offset in GOODS-S \citep{rujopakarn2016,dunlop2017}. 
For source centers in the ZFOURGE catalog, we apply a correction of $-0\farcs086$ in RA and $+0\farcs282$ in Dec with respect to the ALMA image before the above procedure, 
which are calibrated by stars in the Gaia Data Release 1 catalog \citep{gaia2016} within the ASAGAO map. 
{We identify 87 out of 631 ALMA sources that have nearby optical-NIR objects. }
Second, we derive the expected flux density at ALMA Band 6 for the optical-NIR objects from the multi-wavelength spectral energy distribution (SED) fitting with {\sc magphys} \citep{dacunha2008}. 
If the expected flux density is consistent to the total flux density of the ALMA source within $3\sigma$ levels, 
we regard the optical-NIR object as a reliable optical-NIR counterpart of the ALMA source. 
We describe the total flux density measurements and the SED fitting in Section \ref{sec:fp_measure} and Section \ref{sec:our_sample}. 
{We find that 42 out of 87 ALMA sources have reliable optical-NIR counterparts, which we refer to as the ASAGAO catalog.
Note that the number density of the ZFOURGE sources in our ASAGAO map 
provides us the probability of the chance projection ($P$-value; \citealt{downes1986}) in the first step as $\sim$0.03, 
which indicates that $\sim$19 ($=631\times0.03$) out of 87 nearby optical-NIR objects are originated from the chance projection. 
The rejected number in the second step of 45 ($=87-42$) is much larger than 19, 
indicating that the ASAGAO catalog maintains a high purity of the real sources.}
The ASAGAO catalog is summarized in Table \ref{tab:source_catalog}. 
The details of the selection process are presented in Yamaguchi et al. (in prep.). 

\begin{figure*}
\begin{center}
\includegraphics[trim=0cm -0.2cm 0cm 0cm, clip, angle=0,width=0.9\textwidth]{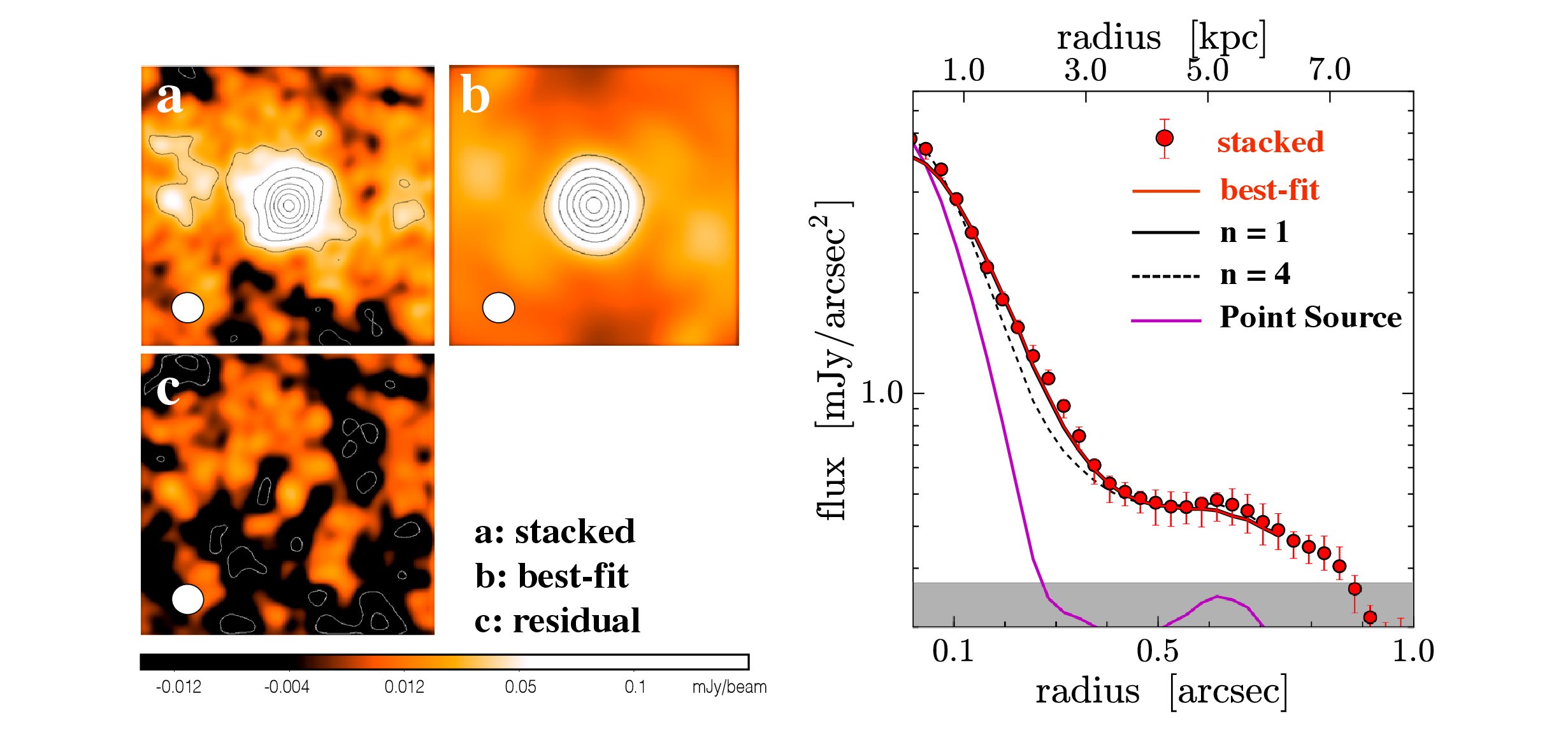}
\vspace{-0.4cm}
 \caption[]{
{\it \bf Left)} {HR (natural-weighted) $1\farcs6$$\times$$1\farcs6$ image} after the visibility-based stacking for the 33 ASAGAO sources at $z=1$$-$3 (a), 
the best-fit S$\acute{\rm e}$rsic profile after beam convolution (b), 
and the residual image (c). 
The black contours denote the 4$-$28$\sigma$ levels with a 4$\sigma$-level step, 
while the white contours indicate the $-2$ and $-4$$\sigma$ levels.  
The synthesized beam ($0\farcs21\times0\farcs17$) is presented at the bottom left in each panel. 
{\it \bf Right)} Radial profile of the surface brightness in the observed frame. 
The red circles and dashed line are the observed values and the best-fit S$\acute{\rm e}$rsic profile for the stacked image, 
where the error-bars are evaluated by the random aperture method. 
The black dashed and solid lines denote the best-fit S$\acute{\rm e}$rsic profiles with the fixed values of $n=1$ and $n=4$, respectively. 
The magenta curve presents the point source profile that corresponds to the synthesized beam of the ALMA observation. 
The gray shade indicates the standard deviation of the pixel values in the stacked image. 
The top axis { depicts} the radius in the kilo-parsec scale for the case that a source resides at $z=2$.
\label{fig:stack_image}}
\end{center}
\end{figure*}

\subsection{Flux and Position Measurement}
\label{sec:fp_measure}

For the flux density and the noise evaluations, 
we use {\sc aegean} \citep{hancock2012} 
which performs a two-dimensional elliptical Gaussian fitting on the image plane. 
If the integrated flux density is smaller than the peak flux density, 
we adopt the peak flux density. 
Otherwise, we adopt the integrated flux density. 
The background and noise are estimated with the BANE package 
that conducts the $3\sigma$ clipping in the signal map, evaluates the standard deviation in a sparse grid of pixels, and then interpolates to make a noise image.
We obtain flux densities of our ASAGAO sources in the range of $\sim$0.3$-$2.3 mJy that are listed in Table \ref{tab:source_catalog}. 
To test the reliability of our flux measurements, we compare the flux measurements of our ASAGAO sources that are also identified with other ALMA Band 6 surveys in GOODS-S \citep{walter2016,dunlop2017,franco2018}. 
We find that the flux measurements are well consistent to other ALMA surveys within the errors. 
The details of the flux and noise measurements and the comparison with other ALMA Band 6 surveys are described in Hatsukade et al. (in prep.).

For the position measurement, 
we conduct a Monte-Carlo (MC) simulation to test whether the LR or HR maps have less positional uncertainty.  
This is because the ALMA sources in the HR map may be more significantly affected by noise fluctuations 
due the reduced surface brightness. 
In the MC simulation, we create 1,000 symmetric artificial sources with a uniform distribution of total flux densities of 0.3$-$1.0 mJy and $R_{\rm e}$ with a range of 0$\farcs$1$-$0$\farcs$3, 
where the $R_{\rm e}$ range is determined from the previous ALMA results \citep[e.g.,][]{ikarashi2015,simpson2015a,hodge2016,fujimoto2017}. 
We then process the artificial sources in the following five steps. 
(I) We inject the artificial source individually into the $uv$-visibilities of the ASAGAO map at a random position, and create a LR map around the artificial source position with a pixel scale of $0\farcs01$. 
(II) We examine whether there is a peak pixel above the 3.5 $\sigma$ level within a radius of $0\farcs5$ from the artificial source center. 
(III) If the peak pixel is detected, we also create a HR map with the same pixel scale as that of the LR map. 
(IV) We determine the source positions in the LR and the HR maps by obtaining the peak pixel positions within a radius of $0\farcs5$ in both the LR and HR maps. 
(V) We obtain the offsets between the peak pixel positions and the artificial source center in both the LR and HR maps. 
We repeat the steps of (I)$-$(V) 1,000 times, changing the injected artificial sources in the step (I).  
Note that we do not perform a profile fitting to determine the source centers in the step (IV) 
because the systematic uncertainties may be large for the faint objects due to low S/N levels. 

The MC simulation results show that the average offset is $\sim$$0\farcs06$ in the LR map and $\sim$$0\farcs10$ in the HR map. 
We thus measure the ASAGAO source centers based on the peak pixel positions in the LR map. 
We discuss the potential effect of the positional uncertainty of $\sim$$0\farcs06$ on the stacking analysis in Section \ref{sec:n-re_est}.

\subsection{Multi-wavelength Properties of Our Sample}
\label{sec:our_sample} 

The details of the physical properties for the ASAGAO sources are presented in Yamaguchi et al (in prep.), 
and here we briefly summarize the general properties of the spectroscopic ($z_{\rm spec}$) and the photometric redshifts ($z_{\rm phot}$), the star-formation rate (SFR), the stellar mass ($M_{\rm star}$), and the IR luminosity ($L_{\rm IR}$) of the ASAGAO sources. 
The values of $z_{\rm spec}$ and $z_{\rm phot}$ values are obtained from the ZFOURGE catalog \citep{straatman2016}. 
The redshift distribution takes the range of $z=0.66$$-$4.36 with median value of $z=1.97$. 
The SFR, $M_{\rm star}$, and $L_{\rm IR}$ values are derived with {\sc magphys} \citep{dacunha2008} based on the rich multi-wavelength data of 47 bands from the rest-frame UV to FIR wavelength in GOODS-S 
that include $Spitzer$/MIPS (Rieke et al. 2004, 24 $\mu$m), $Herschel$/PACS (Poglitsch et al. 2010, 100 and 160 $\mu$m), 
and $Herschel$/SPIRE (Griffin et al. 2010, 250, 350, and 500 $\mu$m), in addition to the ZFOURGE photometry. 
We adopt BC03 templates \citep{bruzual2003} and assume the \cite{chabrier2003} initial mass function and the dust extinction of \cite{charlot2000}. 
The redshift is fixed in the procedure of the SED fitting. 
In the SPIRE bands, we use de-blended SPIRE images based on the 24-$\mu$m source positions as the priors, 
where we perform the de-blending technique in almost the same manner as \cite{liu2018}. 
The details of the de-blending procedure are presented in Wang et al. (in prep.). 
We list $z_{\rm spec}$, $z_{\rm phot}$, SFR, $M_{\rm star}$, and $L_{\rm IR}$ in Table \ref{tab:source_catalog}. 
The ASAGAO sources are generally placed on the massive-end of the main sequence of the SFR$-$$M_{\rm star}$ relation 
with the 16th$-$84th percentiles of SFR $=$ 10.0$-$352 $M_{\odot}$/yr, $\log(M_{\rm star})=$ 10.2$-$11.6 $M_{\odot}$, and $\log(L_{\rm IR})=$ 11.0$-$12.7 $L_{\odot}$. 
The properties of the redshift distribution and the SFR$-$$M_{\rm star}$ relation are consistent with the previous ALMA results for the faint SMGs \citep[e.g.,][]{hatsukade2015b,yamaguchi2016,dunlop2017}, 
which indicates that the ASAGAO sources represent the general population of the faint SMGs. 

\begin{figure}
\begin{center}
\includegraphics[trim=0cm -0.2cm 0cm 0cm, clip, angle=0,width=0.45\textwidth]{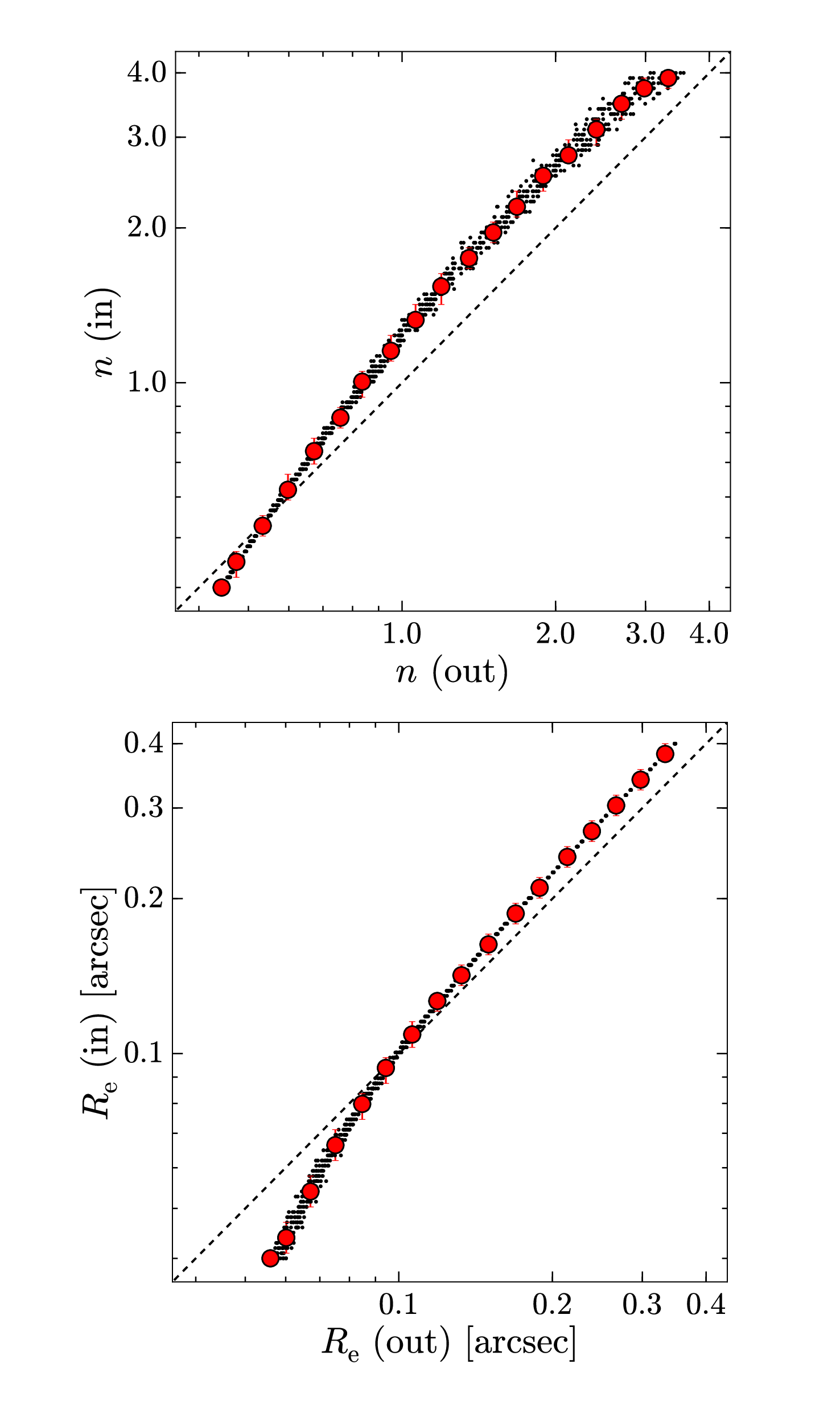}
\vspace{-0.4cm}
 \caption[]{
MC simulation results of the relationship between input and output of S$\acute{\rm e}$rsic index $n$ (top) and effective radius $R_{\rm e}$ (bottom). 
The black dots denote 1,000 model sources.  
In the MC simulation for the $n$ ($R_{\rm e}$) measurement, the input $R_{\rm e}$ ($n$) values are fixed at $R_{\rm e}=0\farcs15$ ($n=1.0$). 
The axis-ratio is also fixed at 0.75. 
The red circles and the error-bars present the median and the 16-84th percentiles of the bins. 
\label{fig:simulation}}
\end{center}
\end{figure}

We also examine the rest-frame optical properties of $n_{\rm opt}$, $R_{\rm e,opt}$, and the morphological classification by utilizing the NIR HST catalogs in GOODS-S. 
For the values of $n_{\rm opt}$ and $R_{\rm e,opt}$, 
the ASAGAO catalog is cross-matched within a radius of $0\farcs5$ with the catalog of \cite{vanderwel2014} that estimate $n$ and $R_{\rm e}$ from the deep HST $J$- and $H$-band images for $H$-band selected sources down to $H\lesssim24.5$. 
Here we take the $n_{\rm opt}$ and $R_{\rm e,opt}$ values from the $J$-band ($H$-band) results for the ASAGAO sources at $z=0.5$$-$1.5 (1.5$-$3.5). 
We identify 21 out of 42 ASAGAO sources whose $n_{\rm opt}$ and $R_{\rm e,opt}$ are reliably (flag $=$ 0) measured in \cite{vanderwel2014}. 
We obtain median values of $n_{\rm opt}=1.5\pm0.01$ and $R_{\rm e,opt}=3.6\pm 0.1$ kpc. 
For the morphological classification, 
we carry out the same cross-matching procedure with the NIR catalog of \cite{kartaltepe2015} 
that complete the visual classification over 50,000 HST/$H$-band selected sources ($H<24.5$). 
The classification is primarily performed with the HST/$H$-band, but the $J$-band image along with 
$V$- and $I$-band ACS images are also used to help the classification. 
There are five morphology classes of disk, spheroid, irregular, compact, and unclassifiable, 
and three additional interaction classes of merger, interaction, and non-interaction 
that are determined by checking close pairs and tidal features. 
We identify 29 out of the 42 ASAGAO sources in the catalog of \cite{kartaltepe2015} with the radius of $0\farcs5$ in the cross-matching. 
A larger search radius of $2\farcs5$ provides us additional one ASAGAO source in the cross-matching, 
which we include in the following analysis not to miss major merger pairs. 
The radius of $2\farcs5$ corresponds to $\sim$20 kpc at $z=2.5$ that is used in \cite{lefevre2000} to search major merger systems. 
The majority of the ASAGAO sources are classified as the disk galaxies, which is consistent with the median $n_{\rm opt}=1.5\pm0.01$. 
We examine the $L_{\rm IR}$ dependence of $n_{\rm opt}$, $R_{\rm e,opt}$, and the morphological classification in Section \ref{sec:result}. 
Note that these rest-frame optical properties for the ASGAO sources are limited by the $H$-band magnitude down to $\sim24.5$. 
However, 34 out of 42 ASAGAO sources have the $H$-band magnitudes over $24.5$, 
indicating that we obtain the general rest-frame optical property from the majority of the ASAGAO sources. 
\begin{figure*}
\begin{center}
\includegraphics[trim=0cm 0cm 0cm -0.4cm, clip, angle=0,width=0.98\textwidth]{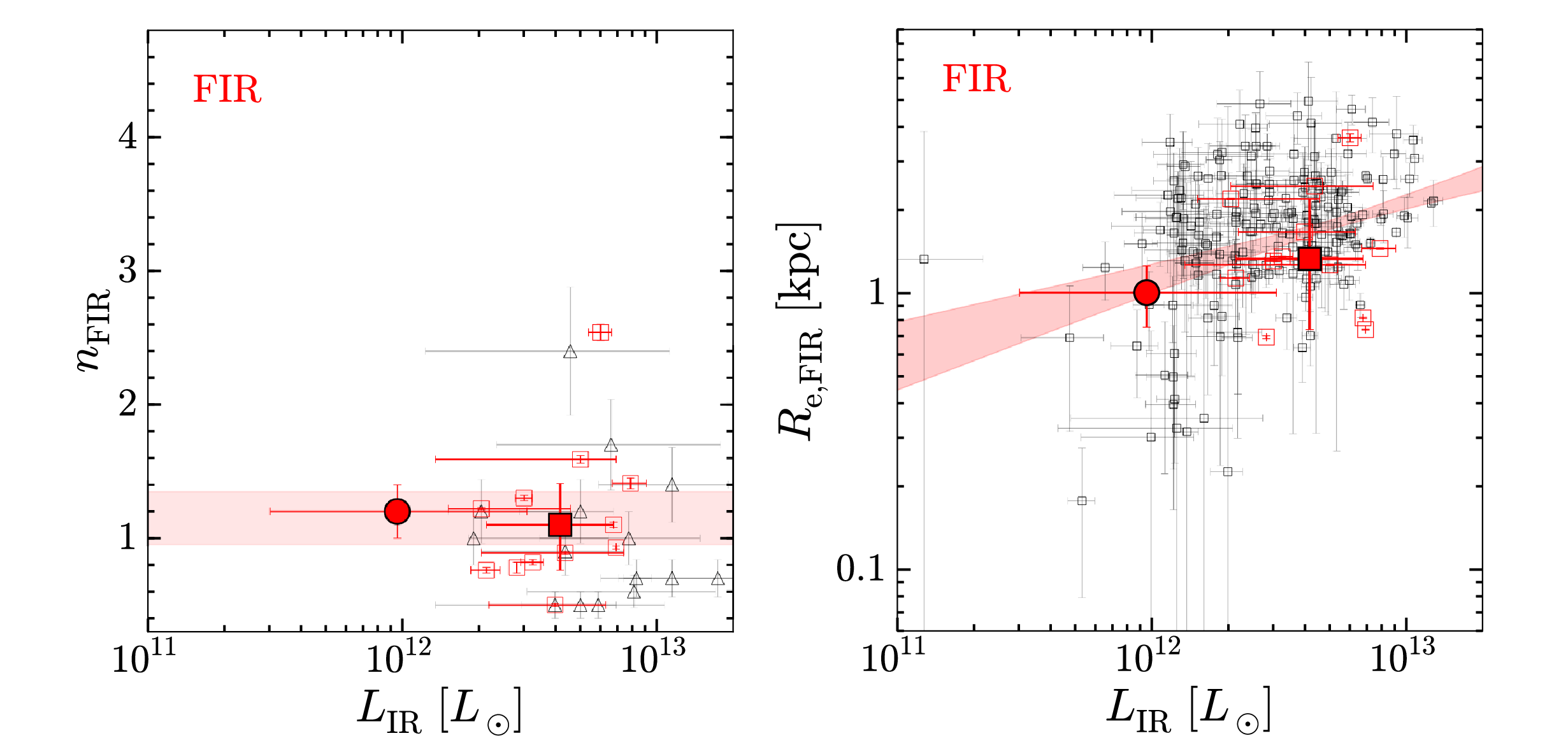}
\vspace{-0.2cm}
 \caption[]{
Rest-frame FIR properties of $n_{\rm FIR}$ (left) and $R_{\rm e,FIR}$ (right) as a function of $L_{\rm IR}$. 
The red filled circle is obtained from the visibility-based stacking for the 33 ASAGAO sources at $z=1$$-$3. 
The $L_{\rm IR}$ error-bar represents the 16th$-$84th percentiles of the $L_{\rm IR}$ distribution for the ASAGAO sources, 
while the $n_{\rm FIR}$ and $R_{\rm e, FIR}$ error-bars are evaluated by the bootstrap method and the MC simulations. 
The red open squares present the additional sample of the 12 individual bright ALMA sources at $z=1$$-$3.
The red filled square indicates the median value of the 12 individual ALMA sources, 
where the error-bars denote the16th$-$84th percentiles of the distribution. 
The black open triangles and squares are the previous ALMA results of \cite{hodge2016} and \cite{fujimoto2017}, respectively. 
{In the right panel, the black open squares are estimated by fixing $n_{\rm FIR}=1$ that we do not present in the left panel.} 
The red shaded regions are the best-estimates of the constant $n_{\rm FIR}$ (left) and the $R_{\rm e,FIR}$$-$$L_{\rm IR}$ relation (right). 
The constant $n_{\rm FIR}$ is estimated from the stacked ASAGAO and the median value of the 12 individual ALMA sources, 
while the best-fit $R_{\rm e,FIR}$$-$$L_{\rm IR}$ relation is obtained from \cite{fujimoto2017}. 
\label{fig:fir_result}}
\end{center}
\end{figure*}

\subsection{Visibility-based Stacking}
\label{stack}
We perform the stacking analysis with {\sc stacker} \citep{lindroos2015} which is a stacking tool on the $uv$-visibility plane for interferometric data. 
{To remove potential effects from the different redshifts such as the redshift evolution of $R_{\rm e, FIR}$ \citep{fujimoto2017} and a slight difference in the rest-frame FIR wavelengths,} 
we use only 33 ASAGAO sources that are located at $z=1-3$ for the stacking.  
The 33 ASAGAO sources span an $L_{\rm IR}$ range of $\sim10^{10}-10^{13}\,L_{\odot}$ with a median value of $10^{11.98}\,L_{\odot}$. 
The center position for the stacking is adjusted to the ASAGAO peak positions that are measured at 1.2-mm wavelength with ALMA Band 6 in Section \ref{sec:fp_measure}. 

Figure \ref{fig:stack_image} shows the HR map after the visibility-based stacking for the 33 ASAGAO sources, 
where the rms noise achieves 8.1 $\mu$Jy/beam. 
The peak flux density of the stacked source shows a 29$\sigma$ significance level that meets the S/N requirement to obtain the reliable fitting results with the S$\acute{\rm e}$rsic profile \citep[e.g.,][]{vanderwel2012,ono2013}. 

\subsection{$n_{\rm FIR}$ and $R_{\rm e,FIR}$ Measurements}
\label{sec:n-re_est}  
We measure $n_{\rm FIR}$ and $R_{\rm e,FIR}$ for the stacked source with {\sc galfit} \citep{peng2010} which is a profile fitting tool on the image plane.  
{We perform the {\sc galfit} task for the $1\farcs6\times1\farcs6$ HR map of the stacked source with the $1\farcs6\times1\farcs6$ synthesized beam image as a point-spread function (PSF). 
We adopt $n=1$, $R_{\rm e}=0\farcs2$, total flux density of 1.0 mJy, axis ratio of 1.0, and position angle of 0 deg as initial parameters, where the sky value is fixed at 0. 
The source center is fixed at the image center. 
We obtain the best-fit results of $n_{\rm FIR}=0.96\pm0.10$ and $R_{\rm e,FIR}=0\farcs11\pm0\farcs02$, where the errors are evaluated by the bootstrap method. 
}
To compare the size measurement on the $uv$-visibility plane, we also use {\sc uvmultifit} \citep{marti2014} that is a profile fitting tool on the $uv$-visibility plane. 
Because one cannot vary the $n_{\rm FIR}$ value in {\sc uvmultifit}, here we measure the $R_{\rm e,FIR}$ value with a fixed value of $n_{\rm FIR}=1$. 
The best-fit $R_{\rm e,FIR}$ is estimated to be $0\farcs10\pm0\farcs03$ that is consistent with the {\sc galfit} result within the error. 

The left panel of Figure \ref{fig:stack_image} presents the best-fit S$\acute{\rm e}$rsic profiles and the residual images for the stacked source.  
The right panel of Figure \ref{fig:stack_image} shows the radial profile of the surface brightness of the stacked source 
with three S$\acute{\rm e}$rsic profiles of our best-fit, $n_{\rm FIR}=1$ (fixed), and  $n_{\rm FIR}=4$ (fixed), 
which demonstrates that the best-fit result is $n_{\rm FIR}\sim1$ instead of $n_{\rm FIR}\sim4$. 

Note that { with the present results} one cannot conclude whether $n_{\rm FIR}\sim1$ or $n_{\rm FIR}\sim4$ is preferable in the range of radius $r>0\farcs4$ ($\sim$3.4 kpc at $z=2$). 
However, it is reasonable to assume that the best-fit result at $r\leq0\farcs4$ represent the major part of the stacked source, 
because the previous studies report that the SMGs have the rest-frame optical effective radius $R_{\rm e,opt}$ of $\sim$3$-$4 kpc \citep[e.g.,][]{targett2011,targett2013,chen2015,fujimoto2017}. 
It should be also noted that we find that there is a slight offset between the stacked and the best-fit profiles near the center ($r\lesssim0\farcs1$). 
In Section  \ref{sec:agn}, we discuss possible origins of this central offset, 
while we mainly focus on the best-fit $n_{\rm FIR}\sim1$ profile at $r\sim0\farcs1$$-$$0\farcs4$ representing the major part of the stacked source profile in this paper. 

Since we find that there potentially exists a positional uncertainty of $\sim0\farcs06$ for the ASAGAO sources in Section 3.2, 
we carry out MC simulations to evaluate how significantly the positional uncertainty affects the $n$ and $R_{\rm e}$ measurements through the stacking process.  
In the MC simulations, first we create 33 model sources that have the same S$\acute{\rm e}$rsic profile with a fixed axis ratio of 0.75.  
We then inject the 33 model sources with random angles and offsets to a two-dimensional plane whose box and grid sizes are the same as the ALMA image used in our $n_{\rm FIR}$ and $R_{\rm e,FIR}$ measurements. 
The offsets of these model sources follow the Gaussian distribution whose average and standard deviation are $0\farcs06$ and $0\farcs01$, respectively. 
We calculate the average radial profile from the 33 model sources after the random injections. 
Finally, we estimate the best-fit S$\acute{\rm e}$rsic profile to obtain $n$ and $R_{\rm e}$ by the minimum chi-square method for the average radial profile. 
We repeat this process 2,000 times, changing the S$\acute{\rm e}$rsic profile of the model sources in the ranges of $n=0.5-4.0$ and $R_{\rm e}=0\farcs04-0\farcs4$. 

Figure \ref{fig:simulation} shows the MC simulation results of the input and output for $n$ and $R_{\rm e}$. 
We find that the output $n$ values are systematically underestimated if the input $n$ values are larger than $\sim0.6$.  
This is because peaky radial profiles with large $n$ values are significantly smoothed by the positional uncertainty through the stacking process {\citep[cf.][]{paulino2018}}. 
We also find that the output $R_{\rm e}$ is deviated from the input $R_{\rm e}$, 
which is also caused by the smoothing effect. 
If the input $R_{\rm e}$ is relatively smaller than the positional uncertainty of $0\farcs06$, 
the smoothed profile makes the output $R_{\rm e}$ larger than the input $R_{\rm e}$. 
On the other hand, if the input $R_{\rm e}$ is relatively large, 
the smoothed profile with small $n$ is difficult to fit the outskirts, 
which makes the output $R_{\rm e}$ slightly smaller than the input $R_{\rm e}$. 
From the MC simulation results, we apply the corrections to our best-fit results of $n_{\rm FIR}$ and $R_{\rm e,FIR}$, and obtain $n=1.2\pm0.2$ and $R_{\rm e,FIR}=0\farcs12\pm0\farcs03$.  
With the median redshift of $z=2.00$ for the ASAGAO sources used in the stacking, 
$R_{\rm e,FIR}$ is estimated to be 1.0$\pm$0.2 kpc.

\subsection{Additional Sample from Archive}
\label{sec:additional_sample} 
To cover a wide range of $L_{\rm IR}$, 
we compile an additional sample of bright SMGs from the ALMA archive. 
We use the ALMA source catalog of \cite{fujimoto2017}, where 1034 ALMA sources (S/N$\geq$5) are identified in ALMA Band 6/7 maps from the ALMA archive. 
We select 12 individual ALMA sources detected with S/N $>$ 15 in the natural-weighted ALMA maps with synthesized beam size of $<0\farcs3$
that have optical-NIR counterparts at $z=$1$-$3. 
In the additional sample, 
we adopt the $L_{\rm IR}$ values estimated in the previous ALMA studies that carry out the SED fitting with optical-FIR bands \citep[e.g.,][]{dacunha2015}. 
Otherwise, we calculate $L_{\rm IR}$ in the same manner as \cite{fujimoto2017} based on an assumption of the modified blackbody with the dust temperature of $T_{\rm d}=35$ K \citep[e.g.,][]{kovacs2006,coppin2008} and the spectral index of $\beta=1.8$ \citep[e.g.,][]{chapin2009,planck2011}. 
The additional sample falls in the $L_{\rm IR}$ range of 10$^{12.3}$$-$$10^{12.9}$ $L_{\odot}$. 
We summarize the properties of the additional sample in Table \ref{tab:additional_catalog}. 

Because the ALMA sources in the additional sample are detected with high S/N levels (S/N $>$ 15), 
we do not perform the stacking for the additional sample. 
The $n_{\rm FIR}$ and $R_{\rm e, FIR}$ values are individually estimated with {\sc galfit} in the same manner as the stacked source, 
where we do not apply the correction due to the stacking described in Section \ref{sec:n-re_est}. 
The median values of $n_{\rm FIR}$ and $R_{\rm e, FIR}$ are estimated to be $n_{\rm FIR}=1.1\pm0.3$ and $R_{\rm e, FIR}=1.3\pm0.8$ kpc, respectively.

\section{Results}
\label{sec:result}

\begin{figure*}
\begin{center}
\includegraphics[trim=-0.2cm -0.2cm 0cm -0.4cm, clip, angle=0,width=0.98\textwidth]{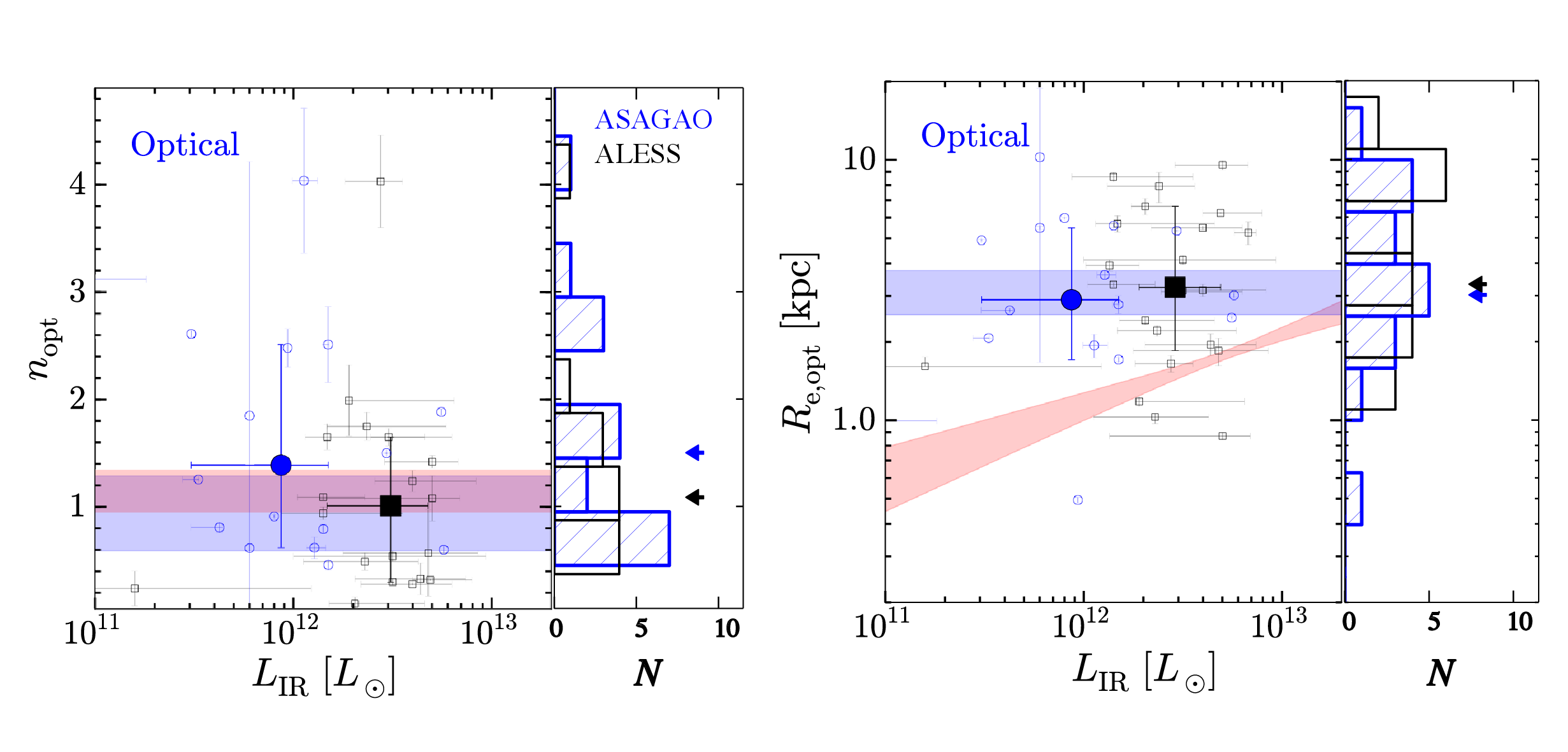}
\vspace{-0.5cm}
 \caption[]{
Rest-frame optical properties of $n_{\rm opt}$ (left) and $R_{\rm e,opt}$ (right) as a function of $L_{\rm IR}$.
The blue open circles are the ASAGAO sources at $z=1-3$ whose $n_{\rm opt}$ and $R_{\rm e,opt}$ values are reliably (flag = 0) measured in \cite{vanderwel2014}. 
The blue filled circles are the median values of the ASAGAO sources.  
The black open squares show the previous ALMA results of \cite{chen2015} for the bright SMGs identified in the ALESS survey. 
The black filled squares are the median values of the ALESS results. 
The error-bars of the blue filled circles and black squares denote the16th$-$84th percentiles of the distribution. 
The blue shaded regions are our best-estimates of constant $n_{\rm opt}$ (left) and $R_{\rm e,opt}$ (right) 
that are derived from the ASAGAO and  ALESS sources. 
The red shaded regions are the same assignment as in Figure \ref{fig:fir_result}. 
The blue and black histograms on the right-side of both panels denote the ASAGAO and ALESS sources, respectively. 
The blue and black arrows present the median values of the ASAGAO and ALESS sources. 
\label{fig:opt_result}}
\end{center}
\end{figure*}

\subsection{FIR Size and Morphology}
\label{sec:fir_result}

Figure \ref{fig:fir_result} shows our stacking results of $n_{\rm FIR}$ and $R_{\rm e,FIR}$ for the ASAGAO as a function of $L_{\rm IR}$. 
To test the $L_{\rm IR}$ dependence of $n_{\rm FIR}$ and $R_{\rm e,FIR}$,  Figure \ref{fig:fir_result} also shows the additional sample of the 12 individual ALMA sources at $z=1-3$. 
For comparison, we also show previous ALMA results from follow-up ALMA high-resolution 870-$\mu$m observations for the bright SMGs identified in An ALMA Survey of Submillimeter Galaxies in the Extended Chandra Deep Field South (ALESS; \citealt{hodge2016}). 

In the left panel of Figure \ref{fig:fir_result}, our stacking result explores the $n_{\rm FIR}$ measurement down to $L_{\rm IR}\sim10^{11} L_{\odot}$. 
We find that our $n_{\rm FIR}$ measurements of both stacking and individual results fall in the $n_{\rm FIR}$ range in the ALESS results of \cite{hodge2016} that derived a median $n_{\rm FIR} = 0.9\pm0.2$. 
The median $n_{\rm FIR}$ value for the individual ALMA sources is ${n_{\rm FIR} =1.1\pm0.3}$, which is consistent with the stacking result from the ASAGAO sources of $n_{\rm FIR}=1.2\pm0.2$, 
suggesting that the dusty star-forming galaxies generally have exponential-disk profiles with $n_{\rm FIR}\sim1$ without a significant dependence on $L_{\rm IR}$. 
Fitting a constant value, we obtain the best-fit $n_{\rm FIR}$ value of $1.2\pm0.2$. 

In the right panel of Figure \ref{fig:fir_result}, 
we find that the our $R_{\rm e,FIR}$ measurements of both stacking and individual results are consistent with the previous ALMA results of \cite{fujimoto2017} 
that {systematically carry out the $R_{\rm e,FIR}$ measurements for the 1034 ALMA sources with a fixed exponential-disk profile of $n_{\rm FIR}=1$}. 
The median $R_{\rm e,FIR}$ value in the individual results is $R_{\rm e,FIR} ={1.3\pm0.8}$ kpc. 
Although this median value is consistent with the stacking result of  $R_{\rm e,FIR} =1.0\pm0.2$ kpc within the $1\sigma$ error, 
this may indicate a trend of a positive correlation between $R_{\rm e,FIR}$ and $L_{\rm IR}$. 
In fact, our $R_{\rm e,FIR}$ measurements show a good agreement with the FIR size$-$luminosity relation of $R_{\rm e,FIR}\propto L_{\rm IR}^{0.28\pm0.07}$ that is derived in \cite{fujimoto2017}.  
Note that we do not estimate the power-law slope value from our $R_{\rm e,FIR}$ measurements. 
This is because it is required to address the selection incompleteness \citep[e.g.,][]{fujimoto2017,kawamata2018} to evaluate the power-law slope value properly, 
which is beyond the scope of this paper. 

\subsection{Optical Size and Morphology}
\label{sec:opt_result}

We also examine the $L_{\rm IR}$ dependence of $n_{\rm opt}$ and $R_{\rm e,opt}$. 
Figure \ref{fig:opt_result} presents the $n_{\rm opt}$ and $R_{\rm e,opt}$ values for the ASAGAO sources at $z=1$$-$3 as a function of $L_{\rm IR}$. 
For comparison, we also show previous results based on follow-up HST/$H$-band observations for the ALESS sources \citep{chen2015}. 
In each panel of Figure \ref{fig:opt_result}, the distributions of the ASAGAO and ALESS sources are shown in the histograms presented in the right side. 

In the histograms of Figure \ref{fig:opt_result}, 
we find that the ASAGAO sources have the similar histograms to the ALESS sources both in $n_{\rm opt}$ and $R_{\rm e,opt}$. 
The Kolmogrov-Smirnov test (KS test) is used to examine whether the histograms of $n_{\rm opt}$ and $R_{\rm e,opt}$ are different statistically. 
The KS test result shows that neither of $n_{\rm opt}$ and $R_{\rm e,opt}$ can rule out the possibility that the histograms are originated from the same parent sample. 
We also perform the Spearman's rank test for the $n_{\rm opt}$ and $R_{\rm e,opt}$ distributions from our ASAGAO and the ALESS sources. 
The Spearman's rank test results show that $n_{\rm opt}$ and $R_{\rm e,opt}$ have the correlation with $L_{\rm IR}$ at the $\sim2\sigma$ and $\sim0\sigma$ levels, respectively. 
The KS and Spearman's rank test results support that neither of $n_{\rm opt}$ and $R_{\rm e,opt}$ depend significantly on $L_{\rm IR}$. 
Although $n_{\rm opt}$ may have a week anti-correlation with $L_{\rm IR}$, 
the statistical significance level is poor with the large scatter in the present result. 
We thus fit constant values to both $n_{\rm opt}$ and $R_{\rm e,opt}$ of the ASAGAO and the ALESS sources, 
and obtain the best-fit values of $n_{\rm opt}=0.9\pm0.3$ and $R_{\rm e,opt}=3.2\pm0.6$ kpc. 

\begin{figure}
\begin{center}
\includegraphics[trim=0.2cm 0cm -2cm 0cm, clip, angle=0,width=0.5\textwidth]{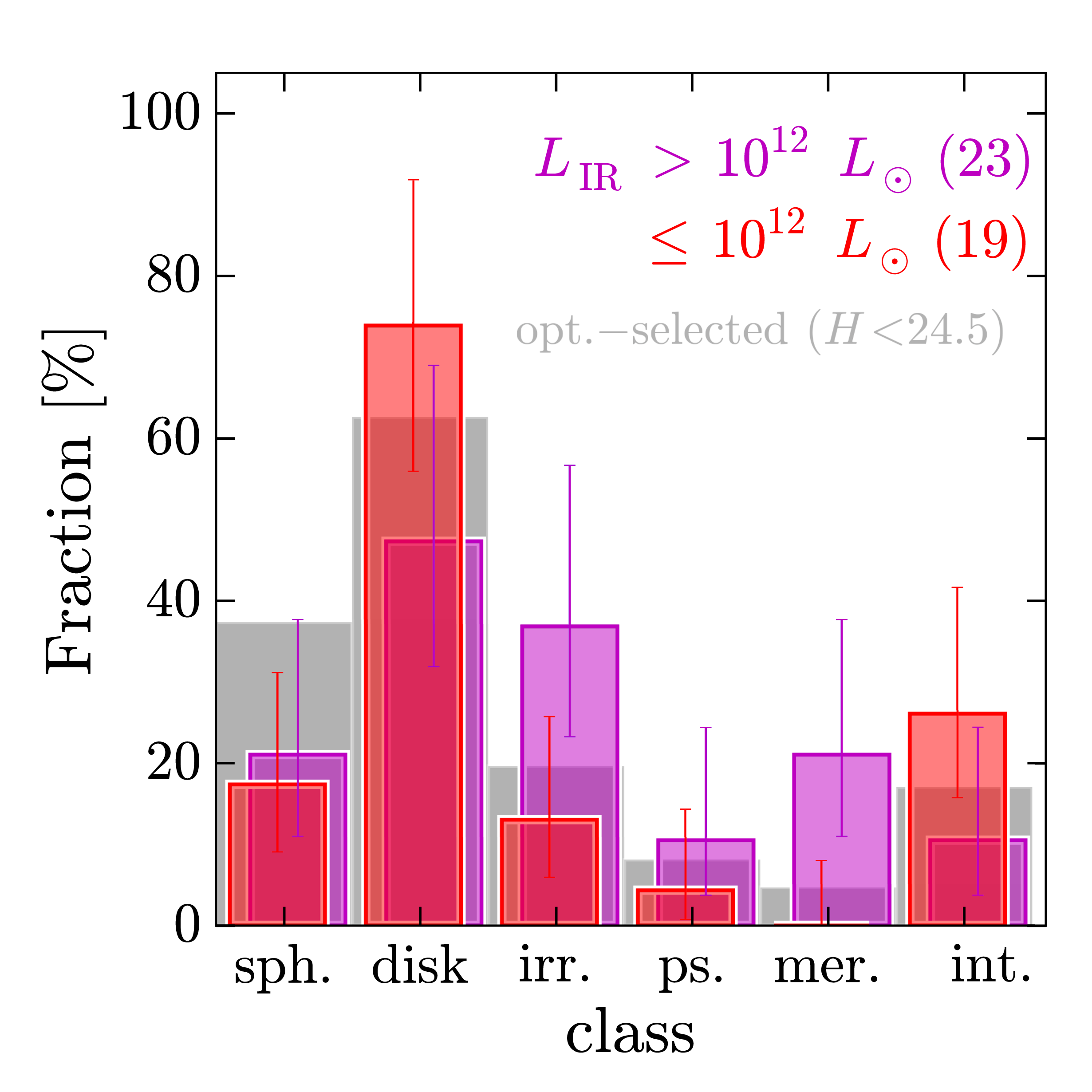}
\vspace{-0.3cm}
 \caption[]{
Histograms of the rest-frame optical morphology with the deep HST/$H$-band images for the ALMA sources at $z=1-3$. 
The morphological classification contains six classes: spheroid (sph.), disk, irregular (irr.), point source (ps.), merger (mer.), and interacting (int.) 
as in \cite{kartaltepe2015}.  
The magenta and red histograms indicate the high-$z$ ULIRG ($L_{\rm IR}>10^{12}\, L_{\odot}$) and LIRG ($L_{\rm IR}\leq10^{12}\, L_{\odot}$) samples, respectively. 
The error-bars denote the Poisson uncertainty presented in \cite{gehrels1986}. 
The number of galaxies in the ULIRG and LIRG samples are 23 and 19, respectively. 
The gray histogram is obtained from the entire sample of the $>$50,000 optically-selected galaxies ($H<24.5$) in the morphology catalog of \cite{kartaltepe2015}.  
Note that the sum of the percentages exceeds 100\% because the morphological classifications are not mutually exclusive. 
The magenta histogram is slightly shifted along the x-axis for clarity. 
\label{fig:opt_morph}}
\end{center}
\end{figure}

\subsection{Comparison between FIR and Optical}
\label{sec:comp_result}

We compare our best-estimates of $n$ and $R_{\rm e}$ in the rest-frame FIR and optical wavelength. 
Figure \ref{fig:opt_result} presents our best-estimates of $n$ and $R_{\rm e}$ in the rest-frame FIR and optical wavelengths with the red and blue shaded regions, respectively. 

In the left panel of Figure \ref{fig:opt_result}, we find that the $n$ value hardly changes between the rest-frame FIR and optical wavelengths with $n_{\rm FIR}\sim n_{\rm opt}\sim1$. 
On the other hand, in the right panel of Figure \ref{fig:opt_result}, we find that $R_{\rm e,FIR}$ is generally smaller than $R_{\rm e,opt}$, 
which is consistent with previous ALMA results that shows that dusty star formation takes place in a compact region \citep[e.g.,][]{simpson2015a,barro2016,hodge2016,tadaki2017a,fujimoto2017}. 
 
Note that our results of $n_{\rm FIR}\sim1$ and small $R_{\rm e,FIR}$ is not the direct evidence of the existence of a compact dusty disk with dynamical rotations. 
However, it is reported that local disk galaxies have inner disk-like structures such as bars, oval disk, and spirals that play an important role to build up the central bulge \citep[e.g.,][]{kormendy2004}. 
This implies that the compact dusty star formation may also contribute to the central bulge formation with the inner disk-like structures. 
In fact, some local galaxy studies show that there are compact dusty disk structures in the stellar disk \citep[e.g.,][]{leeuw2007,holwerda2012}.
In distant galaxies at $z=2.5$, \cite{tadaki2017b} also report molecular gas sizes more compact than the rest-frame optical sizes from two massive star-forming galaxies which contain compact dusty star-forming regions. 
These results may suggest that dusty star formation occurs in a compact dusty disk embedded in a larger stellar disk. 

\subsection{Morphological Classification}
\label{sec:opt_morph}

We also check the $L_{\rm IR}$ dependence of the morphological classification in the rest-frame optical wavelength for the ALMA sources at $z=1-3$. 
To obtain statistically reliable results, we also use the ALMA sources identified in previous ALMA studies in GOODS-S \citep{hodge2013, dunlop2017}, 
and perform the same procedure of the morphological classification as described in Section \ref{sec:our_sample}. 
The whole ALMA sample combining the ASAGAO and the previous ALMA sources span the $L_{\rm IR}$ range of $\sim10^{11}-10^{13}$ $L_{\odot}$,  
where we divide the whole sample into two sub-samples: the high-$z$ ULIRG ($L_{\rm IR}>10^{12}L_{\odot}$) and LIRG ($L_{\rm IR}\leq10^{12}L_{\odot}$) samples. 

Figure \ref{fig:opt_morph} presents the histograms of the morphological classification in the rest-frame optical wavelength for the (U)LIRG samples. 
For comparison, we also show the histogram obtained from the entire sample of the optically-selected galaxies in \cite{kartaltepe2015}. 

In Figure \ref{fig:opt_morph}, 
we find that the LIRG sample contains more disk galaxies and less irregular/merging galaxies than the ULIRG sample. 
The less irregular/merging galaxies in the LIRG sample indicates that 
on-going merger processes are not the dominant trigger of the dusty star formation in the LIRG sample. 
In fact, the fraction of disk galaxies in the LIRG sample is comparable to that of the entire sample of optically-selected galaxies. 
These results imply that dusty star formation in the high-$z$ LIRG sample take place under a secular star-formation mode in the regular disk galaxies. 
On the other hand, the ULIRG sample has significantly larger fractions of the irregular and merging galaxies than the other samples. 
This indicates that the on-going merger process is important in the origin of the dusty star formation in the ULIRG sample. 
Nonetheless, the fraction of disk galaxies is the largest even in the ULIRG sample, 
implying that the dusty star formation in the high-$z$ ULIRGs are triggered also by other mechanisms than the on-going merger process. 
Note that the rest-frame optical morphology classification is limited by the spatial resolution of HST/$H$-band ($\sim0\farcs18$). 
For both LIRG and ULIRG samples, we cannot rule out the possibility that some of the disk galaxies are the mergers, 
where merger pairs are already coalesced and difficult to be resolved.

\section{discussion}
\label{sec:discussion} 

\subsection{Compact Dusty Bulge vs AGN} 
\label{sec:agn}
In this section, we discuss the physical origins of the compact component identified at the stacked source center. 
In the right panel of Figure \ref{fig:stack_image}, 
we find that there is a slight offset between the stacked and the best-fit profiles near the center (radius $\lesssim0\farcs1$). 
The offset is likely to follow the synthesized beam profile, which indicates that this central offset is originated from a very compact component at the center. 
In fact, \cite{hodge2016} also show that some of the bright SMGs have the radial profile that agrees with a Gaussian + point source (PS) profile in the high-resolution ALMA maps. 

There are two possible origins of the central compact component. 
One possibility is a compact dusty bulge with the $n_{\rm FIR}\sim4$ profile which may be related to the bulge formation in the galaxy center. 
The other possibility is an AGN. 
Recent ALMA studies show a radius of $\sim$150-pc scale dust-continuum emission from AGN and surrounding regions in the local galaxies \citep[e.g.][]{izumi2015}. 
In our ALMA HR map, the radius of the synthesized beam (circularized $\sim0\farcs10$) corresponds to $\sim$ 800 pc at $z=2$. 
The central AGN is thus not resolved in our ALMA HR map, and observed as a PS profile. 

To examine which origin is favored to explain the central compact component, 
we conduct a two-component fitting with {\sc galfit} for the $1\farcs6\times1\farcs6$ HR map of the stacked source. 
Here we assume two different profiles of bulge+disk and PS+disk. 
In both cases, the initial parameters for the disk component are the best-fit values derived by the single S$\acute{\rm e}$rsic profile fitting in Section \ref{sec:n-re_est}, 
where $R_{\rm e}$ of the disk component is fixed to obtain stable results. 
The central position of each component is also fixed at the image center. 

In the best-fit results, we find that the minimum chi-square values in the two different profiles are almost the same. 
However, the best-fit result with the bulge+disk profile shows that $R_{\rm e}$ of the bulge component reaches the minimum value in the fitting range of $0\farcs0001$. 
This indicates that the best-fit bulge+disk profile equals to the PS+disk profile. 
From the best-fit result with the PS+disk profile, we obtain the PS component contribution to the total flux density as $1.5\pm0.5\%$.  
Interestingly, a contribution of the AGN to the total flux density 
is estimated to be $\sim0.5\%$ at the $\sim$1-mm flux density 
by fitting the composite star-forming + AGN template for the 16 ALMA sources in \cite{dunlop2017}, 
which shows a good agreement with the value of $1.5\pm0.5\%$ within several factors. 
Moreover, \cite{ueda2018} report a high fraction ($\sim$70\%) of ALMA mm-selected galaxies as X-ray AGNs in the $L_{\rm IR}$ range of 10$^{11.5}$$-$10$^{12.8}\,L_{\odot}$. 
These results may indicate that the compact component is originated from the central AGN. 
Note that one cannot rule out a possibility that there is a PS-like very compact dusty bulge at the center, 
which may be resolved in the future deep ALMA observations with higher angular resolutions.

\begin{figure*}
\begin{center}
\includegraphics[trim=0cm 0cm 0cm 0cm, clip, angle=0,width=1.\textwidth]{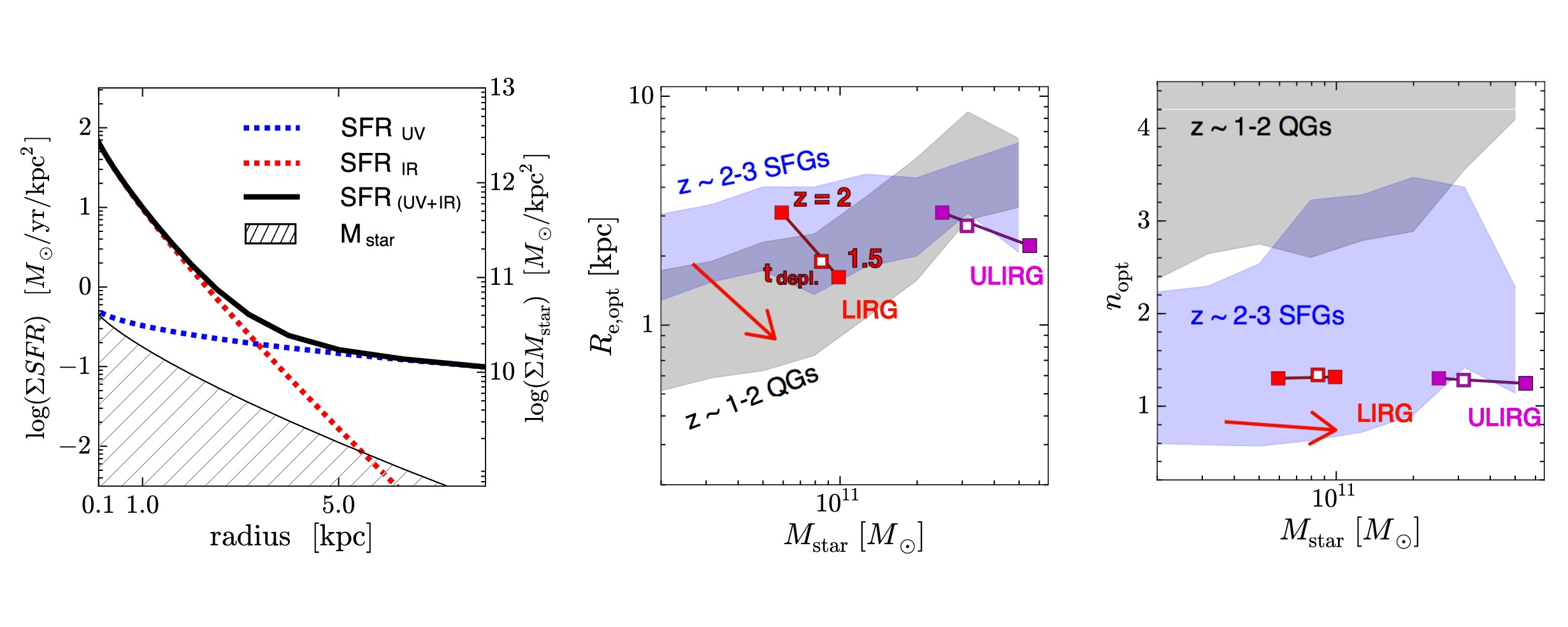}
 \vspace{-0.6cm}
 \caption[]{
{\it \bf Left)} Radial surface density profile of SFR and $M_{\rm star}$ for the high-$z$ LIRGs. 
The red and blue dashed lines represent the radial profile of $\Sigma$SFR$_{\rm IR}$ from our best estimates in the rest-frame FIR wavelength (Section \ref{sec:fir_result}) and $\Sigma$SFR$_{\rm UV}$ from the stacking result with the deep HST/F606W image (see text), respectively. 
The black line indicates $\Sigma$SFR$_{\rm UV+FIR}$ that is obtained by combining the red and blue dashed lines. 
The black shade denotes $\Sigma$$M_{\rm star}$ estimated from our best estimates in the rest-frame optical wavelength (Section \ref{sec:opt_result}). 
{\it \bf Middle)} $R_{\rm e,opt}$ evolution of the high-$z$ (U)LIRGs. 
The red and magenta filled squares show the $R_{\rm e,opt}$ evolutions of high-$z$ ULIRG and LIRGs from $z=2$ to $z=1.5$ under the constant star-formation. 
The red and magenta open squares present the same $R_{\rm e,opt}$ evolutions, but under the constant star-formation until the depletion time scales. 
The red and magenta lines indicate the evolutional tracks of ULIRG and LIRG, respectively. 
The blue and black shaded regions are 16th-84th percentiles of $z=$ 1.5$-$3.0 star-forming (SFGs) and $z=$ 0.5$-$2 quiescent galaxies (QGs) obtained from \cite{shibuya2015}. 
{\it \bf Right)} $n_{\rm e(opt.)}$ evolution of high-z (U)LIRGs. 
The symbols and color assignments are the same as in the middle panel. 
\label{fig:rn_evolv}}
\end{center}
\end{figure*}

\subsection{Evolutions of Size and Morphology}
\label{sec:evolutions}
In this section, we discuss the evolutionary sequence from the high-$z$ dusty star-forming galaxies to the compact quiescent galaxies based on the evolution of the galaxy size and morphology in the rest-frame optical wavelength. 
It has been suggested that there exists an evolutionary connection from high-$z$ dusty star-forming galaxies into local elliptical galaxies via the compact quiescent galaxy (cQG) phase at redshift $z\sim1$$-$2
\citep[e.g.,][]{lilly1999,genzel2003,tacconi2008,hickox2012,toft2014,chen2015,simpson2015a,ikarashi2015,hodge2016,barro2016,fujimoto2017}. 
In the rest-frame optical wavelength, the dusty star-forming galaxies have the exponential-disk profile of $n_{\rm opt}\sim1$ with $R_{\rm e,opt}\sim3$ kpc (Section \ref{sec:opt_result}), 
while the cQGs generally have the spheroidal profile of $n_{\rm opt}\sim4$ with $R_{\rm e,opt}\lesssim1$ kpc \citep[e.g.,][]{vandokkum2008,vanderwel2014,shibuya2015}.  
To confirm the evolutionary connection from the high-$z$ dusty star-forming galaxies to the cQGs, 
it is required to explain the decreasing and increasing trends of $R_{\rm e,opt}$ and $n_{\rm opt}$, respectively, at the same time. 

For the evolutionary mechanisms, there have been several arguments in the literature, 
such as mergers of gas-rich galaxies \citep[e.g.,][]{tacconi2008}, 
inside-out growth of compact progenitors \citep[e.g.,][]{nelson2014,wellons2015}, 
and ``compaction'' of the gas in star-forming galaxies due to disk instabilities \citep{dekel2014}. 
Here we alternatively investigate the evolutionary mechanisms by evaluating the contribution of the pure star-forming activity in the dusty star-forming galaxies to the $R_{\rm e,opt}$ and $n_{\rm opt}$ evolutions. 
Note that intense star-formation in the dusty starbursts galaxies is known to have high SFRs even up to the order of $\sim$1000 $M_{\odot}$/yr, 
but have short depletion times of $\sim$100$-$200 Myr \citep[e.g.,][]{carilli2013}. 
On the other hand, a secure star-formation in normal star-forming galaxies has a longer depletion time of $\sim$1 Gyr than the intense star-formation \citep[e.g.,][]{tacconi2010}. 
This may indicate that the secular star-formation contributes to the majority of the stars instead of the intense star-formation \citep[e.g.,][]{elbaz2011,rujopakarn2011,wuyts2011,schreiber2016}. 
We thus examine the $R_{\rm e,opt}$ and $n_{\rm opt}$ evolutions for not only the high-$z$ ULIRGs ($L_{\rm IR}>10^{12}L_{\odot}$), 
but also LIRGs ($L_{\rm IR}=10^{11}$$-$$10^{12}L_{\odot}$) 
that are regarded as the galaxies under the intense and the secular star-formation modes, respectively. 

To obtain the general physical properties for the high-$z$ (U)LIRGs, 
we divide the 33 ASAGAO sources at $z=1-3$ into two $L_{\rm IR}$ samples of $L_{\rm IR}\leq10^{12}$ and $>10^{12}\,L_{\odot}$ 
that we assume as the representative samples of the high-$z$ LIRGs and ULIRGs, respectively. 
For the general physical property in the rest-frame FIR and optical wavelengths, 
we evaluate the average $L_{\rm IR}$ and $M_{\rm star}$ individually for the high-$z$ LIRG and ULIRG samples. 
Based on the average $L_{\rm IR}$, we evaluate the average S$\acute{\rm e}$rsic profiles in the rest-frame FIR and optical wavelengths from our best-fit results as a function of $L_{\rm IR}$ derived in Section \ref{sec:result}. 
For the general physical property in the rest-frame UV wavelength, 
we carry out the image-based stacking for the 33 ASAGAO sources, utilizing the deep HST/F606W image in GOODS-S. 
Here the source centers for the stacking are adjusted to the ASAGAO source centers that are measured in ALMA Band 6, 
because the rest-frame FIR emission represents the massive galaxy centers rather than the rest-frame UV emission \citep[e.g.,][]{barro2016,tadaki2017a}. 
From the best-fit S${\acute{\rm e}}$rsic profile for the stacked HST/F606W image with {\sc galfit}, 
we obtain the average values of rest-frame UV S$\acute{\rm e}$rsic index $n_{\rm UV}$, effective radius $R_{\rm e,UV}$, and the un-obscured UV luminosity ($L_{\rm UV}$). 
We also calculate the obscured (SFR$_{\rm IR}$) and un-obscured SFR (SFR$_{\rm UV}$) from the average $L_{\rm IR}$ and $L_{\rm UV}$ with the equation in \cite{straatman2016} of
\begin{eqnarray}
{\rm SFR}_{\rm IR} &=& 1.09 \times 10^{-10}\, L_{\rm IR} (L_{\odot}), \\
{\rm SFR}_{\rm UV}  &=& 2.4  \times 10^{-10}\, L_{\rm UV} (L_{\odot}). 
\end{eqnarray}

The left panel of Figure \ref{fig:rn_evolv} shows radial surface density profiles of $M_{\rm star}$ ($\Sigma$$M_{\rm star}$), SFR$_{\rm IR}$ ($\Sigma$SFR$_{\rm IR}$), and SFR$_{\rm UV}$ ($\Sigma$SFR$_{\rm UV}$) for the high-$z$ LIRGs. 
Here we derive the radial profile of $\Sigma$$M_{\rm star}$ from the rest-frame optical S$\acute{\rm e}$rsic profile with HST/$H$-band in Section \ref{sec:opt_result}, 
assuming that the source centers are the same in the rest-frame optical and FIR wavelengths. 
The radial surface density profile of the total SFR ($\Sigma$SFR$_{\rm UV+FIR}$) is also presented by combining $\Sigma$SFR$_{\rm UV}$ and $\Sigma$SFR$_{\rm IR}$. 
In the radial surface density profiles, we find that $\Sigma$SFR$_{\rm IR}$ dominates over $\Sigma$SFR$_{\rm UV}$ in the central star-forming region both for the high-$z$ LRIGs and ULIRGs. 

Assuming a constant SFR during a time scale $\Delta t$ and no radial motions of the stars, 
we derive the $R_{\rm e,opt}$ and $n_{\rm opt}$ evolutions for the high-$z$ ULRGs and LIRGs that are estimated from the sum of the $\Sigma M_{\rm star}$ and $\Sigma$SFR$_{\rm UV+FIR}$$\times$$\Delta t$. 
Here we adopt two different time scales. 
First is the depletion time, where we obtain the gas mass ($M_{\rm gas}$) from the gas fraction ($f_{\rm gas}$) as a function of $M_{\rm star}$ \citep[e.g.,][]{tacconi2013,mirka2015}. 
The depletion times are estimated to be $\sim$ 640 and 210 ${\rm Myr}$ for the high-$z$ LIRGs and ULIRGs, respectively. 
Second is the cosmic time of $\sim1.0$ ${\rm Gyr}$ from $z=2$ to 1.5, 
where we assume a case that the star-formation continues in a longer time than the depletion time due to the gas supply from inflows. 
In fact, the recent bathtub model results show that $f_{\rm gas}$ does not significantly change down to $z\sim1$ 
in the balance of the gas inflow/outflow and the gas consumption by the star-formation \citep[e.g.,][]{dekel2014}. 

The middle and right panels of Figure \ref{fig:rn_evolv} show the results of the $R_{\rm e,opt}$ and $n_{\rm opt}$ evolutions for the high-$z$ ULIRGs and LIRGs in our model. 
For comparison, the distributions of $z\sim2-3$ star-forming and $z\sim1$$-$2 quiescent galaxies are also presented with the blue and black shaded regions, respectively. 
In the middle panel of Figure \ref{fig:rn_evolv}, we find that the decreasing trend of $R_{\rm e,opt}$ is reproduced well both for the high-$z$ ULIRGs and LIRGs. 
On the other hand, we find that the increasing trend of $n_{\rm opt}$ is not reproduced. 
Because our model assumes no radial motions of the stars, 
this result implies that other mechanisms are required to form the spheroidal profile of $n_{\rm opt}\sim4$ such as dynamical dissipation from the star-forming to quiescent galaxies. 
In fact, a recent spectroscopic study has discovered a cQG at $z=2.1478$ that turns out to be a fast-spinning, rotationally supported galaxy with $n_{\rm opt}=1.01^{+0.12}_{-0.06}$ \citep{toft2017}. 
Unless the angular momentum in a post star-forming galaxy is lost by any dynamical dissipating events, 
it is reasonable that the stellar distribution maintains the exponential-disk profile of $n_{\rm opt}\sim1$. 
Our model and the recent observational results suggest another step for the star-forming galaxies to form the spheroidal stellar distribution after quenching. 
Note that the initial $M_{\rm star}$ values of the high-$z$ (U)LIRGs in our model are relatively massive, compared with the main-sequence of the star-forming galaxies at $z\sim2$. 
It may indicate that less massive, normal/dusty star-forming galaxies at higher redhshifts are more related to the origin of  the cQGs rather than $z\sim$2 dusty star-forming galaxies. 
To fully understand the origin of the cQGs, it is also required to examine the evolutions of size and morphology from the normal and dusty star-forming galaxies at higher redshifts in the future deep observations.

\section{Summary}
\label{sec:summary}
In this paper, we study the S$\acute{\rm e}$rsic index $n$ and the effective radius $R_{\rm e}$ in the rest-frame FIR wavelength, $n_{\rm FIR}$ and $R_{\rm e,FIR}$, 
via the visibility-based stacking method for the ALMA sources identified in the ASAGAO survey. 
The high-resolution ($\sim0\farcs19$) ALMA 1.2-mm imaging of our ASAGAO survey covers the 26 arcmin$^{2}$ area in GOODS-S imaged with WFC3/IR on HST, 
which also enables us to obtain $n$ and $R_{\rm e}$ in the rest-frame optical wavelength, $n_{\rm opt}$ and $R_{\rm e,opt}$.   
In conjunction with the individual ALMA sources from the archive and previous ALMA results, 
we examine the $L_{\rm IR}$ dependence of the morphology of the dusty star-forming galaxies in the rest-frame FIR and optical wavelengths in the wide $L_{\rm IR}$ range of $\sim10^{11}-10^{13}\,L_{\odot}$. 
We then discuss the evolutionary connections from the high-$z$ dusty star-forming galaxies to the compact quiescent galaxies. 
The major findings of this paper are summarized below.
\begin{enumerate}
\item
The visibility-based stacking of the 33 ASAGAO sources at $z=1-3$ produces a 29$\sigma$ level detection 
which enables us to have a reliable S$\acute{\rm e}$rsic profile fit. 
Evaluating the positional uncertainty and the smoothing effect by realistic Monte-Carlo simulations, 
we obtain the best-estimates of $n_{\rm FIR}=1.2\pm0.2$ and $R_{\rm e,FIR}=1.0\pm0.2$ kpc. 

\item
With individual measurements of 12 ALMA sources from the archive, 
we evaluate the $L_{\rm IR}$ dependence of $n_{\rm FIR}$ and $R_{\rm e,FIR}$ in the $L_{\rm IR}$ range of $\sim10^{11}-10^{13}\,L_{\odot}$. 
We find that the $n_{\rm FIR}$ measurements hardly change as a function of $L_{\rm IR}$ with $n_{\rm FIR}=1.2\pm0.2$. 
On the other hand, the $R_{\rm e,FIR}$ measurements shows a good agreement with the positive power-law correlation between $R_{\rm e,FIR}$$-$$L_{\rm IR}$ identified in previous ALMA studies. 
\item 
The distributions of $n_{\rm opt}$ and $R_{\rm e,opt}$ show no significant difference between our ASAGAO sources and the bright SMGs identified in the ALESS survey. 
The statistical test results suggest that neither $n_{\rm opt}$ and $R_{\rm e,opt}$ depend strongly on $L_{\rm IR}$. 
We obtain the best-estimates of $n_{\rm opt}=0.9\pm0.3$ and $R_{\rm e,opt}=3.2\pm0.6$ kpc in the $L_{\rm IR}$ range of $\sim10^{11}-10^{13}\,L_{\odot}$. 
\item 
Comparing our rest-frame FIR and optical results, 
we find that $n$ takes the common value of $\sim1$ both in the rest-frame FIR and optical wavelengths. 
We also find that $R_{\rm e,FIR}<R_{\rm e,opt}$, which is consistent with the previous ALMA studies. 
These results indicate that the dusty star formation takes place in a compact {\bf disk-like structure} which is embedded in a larger stellar disk.  
\item 
The detail morphological study in the rest-frame optical wavelength for the $z\sim1-3$ ALMA sources 
suggests that on-going merger processes are likely important in high-$z$ ULIRGs, but not in LIRGs. 
However, the fraction of disk galaxies is the largest even in the high-$z$ ULIRGs, 
implying that the dusty star formation is triggered not only by the on-going merger process even in the ULIRGs at $z=$ 1$-$3. 

\item
In the HR map of the stacked ASAGAO source, 
we find that there is a compact component at the center. 
The two-component fitting results show that the stacked profile is explained by a point source (PS)+disk profile rather than a bulge+disk profile. 
The PS component is likely to be originated by the central AGN, 
while one cannot rule out a possibility that there is a PS-like very compact dusty bulge at the center.  

\item
The best-fit S$\acute{\rm e}$rsic profiles in the rest-frame FIR+UV and optical 
provides us the radial surface density profile of SFR$_{\rm UV+FIR}$ and $M_{\rm star}$, respectively. 
Assuming the constant star-formation, our simple model reproduces the decreasing trend of $R_{\rm e,opt}$ from $z\sim2$ dusty star-forming galaxies to $z\sim1-2$ compact quiescent galaxies. 
However, the increasing trend of $n_{\rm opt}$ is not reproduced, 
where other mechanism(s) are required such as kinematic dissipation. 
\end{enumerate}

We are grateful to Ivan Marti-Vidal and the Nordic ALMA Regional Center for providing us helpful CASA software tools and advice on analyzing the data. 
We appreciate Yoshiaki Ono, Takashi Kojima, and Yuichi Harikane for useful comments and suggestions. 
We are indebted for the support of the staff at the ALMA Regional Center. 
This paper makes use of the following ALMA data: ADS/JAO. ALMA \#2015.1.00098.S, \#2012.1.00307.S, \#2013.1.00576.S, and \#2013.1.00781.S. 
ALMA is a partnership of the ESO (representing its member states), 
NSF (USA) and NINS (Japan), together with NRC (Canada), MOST and ASIAA (Taiwan), and KASI (Republic of Korea), 
in cooperation with the Republic of Chile. 
The Joint ALMA Observatory is operated by the ESO, AUI/NRAO, and NAOJ. 
This study is supported by Grant-in-Aid for JSPS Research Fellow, 
the JSPS Grant-in-Aid for Scientific Research (S) JP17H06130, 
and the NAOJ ALMA Scientific Research Grant Number 2017-06B.
S.I. gratitudes for the funding from JSPS Grant-in-Aid for Young Scientists (B), No. 17K17677, and from CREST, JST. 
T.O. has been supported by MEXT as ‘Priority Issue on Post-K computer’ (Elucidation 
of the Fundamental Laws and Evolution of the Universe) and JICFuS. 
W.R. is supported by JSPS KAKENHI Grant Number JP15K17604 and the Thailand Research Fund/Office of the Higher Education Commission Grant Number MRG6080294. 
HU is supported by JSPS KAKENHI Grant Number JP17K14252. 
S.T. and C.G.G. acknowledge support from the ERC Consolidator Grant funding scheme (project ConTExt, grant number 648179), 
Kavli IPMU is supported by World Premier International Research Center Initiative (WPI), MEXT, Japan.
The Cosmic Dawn Center is funded by the Danish National Research Foundation. 
S.F. was supported by the ALMA Japan Research Grant
of NAOJ Chile Observatory, NAOJ-ALMA-179.

\bibliographystyle{apj}
\bibliography{apj-jour,reference}
\end{document}